\documentclass[acmlarge, nonacm]{acmart}

\usepackage{colortbl}

\usepackage{graphicx} 
\usepackage{colortbl} 
\renewcommand\footnotetextcopyrightpermission[1]{}
\usepackage{amsmath, amssymb} 
\usepackage{algorithm} 
\usepackage{algorithmic} 
\usepackage{graphicx} 
\usepackage{subcaption} 
\usepackage{xcolor} 
\usepackage{multirow} 
\usepackage{array} 
\usepackage{longtable} 
\usepackage{rotating} 
\usepackage{longtable}
\usepackage{url} 
\usepackage{textcomp} 
\usepackage{booktabs} 
\usepackage{verbatim} 

\newcolumntype{C}[1]{>{\centering\arraybackslash}p{#1}} 
\newcolumntype{M}[1]{>{\centering\arraybackslash}m{#1}} 

\definecolor{IEEE_ACCESS_BLUE}{RGB}{43, 171, 225} 
\definecolor{LightGray}{gray}{0.9} 

\usepackage{fontawesome} 

\def\BibTeX{{\rm B\kern-.05em{\sc i\kern-.025em b}\kern-.08em%
    T\kern-.1667em\lower.7ex\hbox{E}\kern-.125emX}}
\setcopyright{none}
\usepackage{xcolor}
\usepackage{colortbl}
\usepackage{pifont}
\usepackage{makecell}
\usepackage{wasysym}
\newcommand{\halfcirc}{\LEFTcircle}

\begin{document}

\title{LLM-Powered Agentic AI for 5G/6G Networks: A Tutorial and Survey on Architectures, Protocols, and Standardization}


\author{Mazene Ameur}
\email{mazene.ameur@eurecom.fr}
\affiliation{%
  \institution{EURECOM}
  \department{Communication Systems Department}
  \city{Sophia Antipolis}
  \country{France}
}

\author{Abdelkader Mekrache}
\email{abdelkader@simula.no}
\affiliation{%
  \institution{Simula Metropolitan}
  \department{CRNA}
  \city{Oslo}
  \country{Norway}
}

\author{Bouziane Brik}
\email{bbrik@sharjah.ac.ae}
\affiliation{%
  \institution{University of Sharjah}
  \department{Computer Science Department}
  \city{Sharjah}
  \country{UAE}
}

\author{Adlen Ksentini}
\email{adlen.ksentini@eurecom.fr}
\affiliation{%
  \institution{EURECOM}
  \department{Communication Systems Department}
  \city{Sophia Antipolis}
  \country{France}
}

\renewcommand{\shortauthors}{Mazene et al.}

\begin{abstract}

Agentic Artificial Intelligence (AI), enabled by Large Language Models, marks a shift from rule-based automation toward autonomous, goal-driven control of Next-Generation Networks (NGNs). Existing surveys treat the two domains in isolation, leaving protocol integration, evaluation, and standardization alignment underexplored. To address this gap, a two-part tutorial-and-survey is presented. Part I formalises the control, management, and AI-native planes of 5G and 6G. It then covers the foundations of agentic systems: reasoning, planning, tool use, multi-agent coordination, and evaluation. Part II maps agentic capabilities onto 5G/6G control surfaces, standardization, and major 6G initiatives. Finally, it identifies open challenges shaping autonomous telecommunications.

\end{abstract}

\ccsdesc[500]{General and reference~Surveys and overviews}
\ccsdesc[500]{Networks~Network design principles}
\ccsdesc[500]{Computing methodologies~Artificial intelligence}


\keywords{Agentic AI, Large Language Models, 5G, 6G, Next-Generation Networks, Network Management, AI-native Networks}


\maketitle

\section{Introduction}
\subsection{Context and Motivation}
 The telecommunications sector is at a critical inflection point shaped by the maturation of Fifth Generation (5G) networks and the early architectural formulation of Sixth Generation (6G) systems \cite{sec1.ref1,sec1.ref2}. Modern network infrastructures are increasingly defined by extreme heterogeneity, ultra-dense deployments, and stringent performance requirements across latency, reliability, and Quality of Service (QoS) \cite{sec1.ref2.5, sec1.ref5.5}.

As the vision for beyond-5G and 6G networks matures, the community increasingly converges on architectures that demand unprecedented levels of automation, closed-loop intelligence, and operational autonomy across heterogeneous infrastructures \cite{sec1.ref4}. Within this trajectory, leading standardization bodies such as 3rd Generation Partnership Project (3GPP\footnote{https://www.3gpp.org/}), European Telecommunications Standards Institute (ETSI\footnote{https://www.etsi.org/}), and TeleManagement (TM) Forum\footnote{https://www.tmforum.org/}, together with the global research community, consistently highlight the need for cohesive frameworks that translate high-level service expectations into actionable, verifiable, and continuously optimized network behaviors at scale \cite{sec1.ref5}.
In this context, Zero-Touch network and Service Management (ZSM) \cite{sec6.b5} and Intent-Based Networking (IBN) emerge as pivotal enablers, not as isolated features but as integral pillars that support the end-to-end autonomy envisioned for Next-Generation Networks (NGNs) \cite{mekrache2024intent}.



Against this backdrop, the emergence of Agentic Artificial Intelligence (AI), enabled by recent advances in Large Language Models (LLMs) and reasoning-centric architectures, marks a fundamental transition from reactive, rule-based automation to proactive, goal-driven intelligence \cite{sec1.ref6}. Unlike traditional AI solutions deployed as isolated optimization components, agentic systems exhibit autonomy, contextual awareness, and coordinated behavior, allowing them to reason over network states, translate high-level intents into actionable policies, and operate across heterogeneous infrastructure layers \cite{sec2.b4}. This capability enables continuous adaptation, self-directed decision-making, and scalable intelligence at the network level, effectively redefining the operational paradigm of NGNs \cite{sec2.b8}.

\subsection{Related Surveys, Gaps, Scope and Objectives}

\textbf{Selection Process.}  The body of related literature considered in this survey was collected through systematic searches on Google Scholar, complemented by a targeted examination of existing surveys with narrower scopes. The search strategy employed representative keywords such as ``Agentic AI,'' ``LLM agents,'' and ``Large Language Model agents,'' augmented with networking-specific terms including ``Next Generation Network,'' ``5G,'' and ``6G networks,'' to ensure comprehensive coverage of potentially relevant studies.
From the initial pool of collected publications, a rigorous pre-selection process was conducted to exclude out-of-scope studies, non-survey papers, and sources lacking sufficient academic rigor. This process resulted in a refined and high-quality set of surveys that are closely aligned with the objectives of this work. A detailed comparative analysis of this curated set enabled the identification of key limitations, unresolved challenges, and underexplored research directions in the existing literature. These observations directly motivated the development of the present survey, which aims to address a critical gap in the intersection of Agentic AI and NGNs, as discussed in this subsection.

\begin{table*}[!t]
\centering
\caption{Summary of related surveys on LLM-based and Agentic AI for networks.}
\scriptsize
\renewcommand{\arraystretch}{1.0}
\setlength{\tabcolsep}{3pt}
\begin{tabular}{p{0.6cm}p{5cm}ccccccc}
\hline
\rowcolor{lightgray}
\textbf{Ref.} &
\textbf{Key Contributions} &
\textbf{LLMs} &
\makecell{\textbf{Agentic} \\ \textbf{Concepts}} &
\makecell{\textbf{Agentic} \\ \textbf{Protocols}} &
\makecell{\textbf{Agentic} \\ \textbf{Evaluation}} &
\makecell{\textbf{Agentic} \\ \textbf{Benchmarks}} &
\makecell{\textbf{Next Gen.} \\ \textbf{Networks}} &
\makecell{\textbf{Standards} \\ \textbf{\& Projects}} \\
\hline\hline

\rowcolor{gray!10}
\cite{sec2.b1} & LLM survey for 6G: architectures, communication applications, agentic concepts.
& \halfcirc & \halfcirc & \ding{55} & \checkmark & \ding{55} & \checkmark & \ding{55} \\
\hline

\cite{sec2.b2} & Agent communication security: lifecycles, MCP/A2A protocols, defense mechanisms.
& \checkmark & \checkmark & \checkmark & \halfcirc & \ding{55} & \ding{55} & \ding{55} \\
\hline

\rowcolor{gray!10}
\cite{sec2.b3} & LLM-enabled network ops: fault diagnosis, causal inference, optimization.
& \checkmark & \ding{55} & \ding{55} & \ding{55} & \ding{55} & \checkmark & \ding{55} \\
\hline

\cite{sec2.b4} & Agentic AI survey: autonomy, memory, goal-driven reasoning, ethics.
& \halfcirc & \checkmark & \ding{55} & \ding{55} & \ding{55} & \halfcirc & \ding{55} \\
\hline

\rowcolor{gray!10}
\cite{sec2.b5} & AGI-native wireless: cognitive telecom brain for autonomous network control.
& \halfcirc & \halfcirc & \ding{55} & \ding{55} & \halfcirc & \checkmark & \ding{55} \\
\hline

\cite{sec2.b6} & GenAI survey for 6G: model families, learning paradigms, security applications.
& \halfcirc & \ding{55} & \ding{55} & \ding{55} & \ding{55} & \checkmark & \ding{55} \\
\hline

\rowcolor{gray!10}
\cite{sec2.b7} & Edge General Intelligence: distributed agents, world models, foundation models.
& \halfcirc & \checkmark & \ding{55} & \halfcirc & \ding{55} & \checkmark & \ding{55} \\
\hline

\cite{sec2.b8} & AI agents in 6G: use cases, collaborative architectures, system challenges.
& \halfcirc & \checkmark & \ding{55} & \ding{55} & \ding{55} & \checkmark & \ding{55} \\
\hline

\rowcolor{gray!10}
\cite{sec2.b9} & LLM agent evaluation: taxonomies for objectives, processes, reliability.
& \halfcirc & \checkmark & \ding{55} & \checkmark & \checkmark & \ding{55} & \ding{55} \\
\hline

\cite{sec2.b10} & LLM chatbots vs.\ agents: evaluation benchmarks and assessment dimensions.
& \checkmark & \checkmark & \ding{55} & \checkmark & \checkmark & \ding{55} & \ding{55} \\
\hline

\rowcolor{gray!10}
\cite{sec2.b11} & LLM and Agentic AI for 6G: design principles, multi-agent frameworks.
& \checkmark & \checkmark & \checkmark & \checkmark & \ding{55} & \checkmark & \ding{55} \\
\hline

\cite{sec2.b13} & AI agent study: single- vs.\ multi-agent designs, coordination mechanisms.
& \halfcirc & \checkmark & \ding{55} & \checkmark & \checkmark & \ding{55} & \ding{55} \\
\hline

\rowcolor{gray!10}
\cite{sec2.b14} & AI Agents vs.\ Agentic AI: taxonomies, hallucination, coordination failures.
& \halfcirc & \checkmark & \ding{55} & \checkmark & \halfcirc & \ding{55} & \ding{55} \\
\hline

\cite{sec2.b15} & LLM/agent benchmarks: evaluation frameworks, protocols, security vulnerabilities.
& \halfcirc & \checkmark & \checkmark & \checkmark & \checkmark & \halfcirc & \ding{55} \\
\hline

\rowcolor{gray!10}
\cite{sec2.b16} & Agentic AI for edge networks: agentification frameworks, edge intelligence.
& \halfcirc & \checkmark & \halfcirc & \halfcirc & \halfcirc & \halfcirc & \ding{55} \\
\hline

\textbf{Ours} & Full survey: Agentic AI for NGNs, protocols, benchmarks, standardization.
& \checkmark & \checkmark & \checkmark & \checkmark & \checkmark & \checkmark & \checkmark \\
\hline

\end{tabular}
\\[4pt]
\footnotesize{Legend: \checkmark = Addressed \quad \halfcirc = Partially Addressed \quad \ding{55} = Not Addressed}
\label{tab:related_surveys}
\end{table*}

\textbf{Related Surveys.} The intersection of LLMs and 5G/6G Networks has only recently attracted survey-level attention. Existing works primarily examine isolated aspects of AI-enabled networking and reflect an early transition from learning-based automation toward foundation model-driven designs. As summarized in Table \ref{tab:related_surveys}, the available surveys remain limited in number and fragmented in scope, with gaps in agentic depth, architectural coverage, protocol realization, evaluation rigor, and alignment with standardization efforts. This fragmentation prevents a unified system-level view of LLM-powered Agentic AI for 5G evolution and emerging 6G networks.

A first group of surveys examines the adoption of LLMs and generative AI in 6G architectures \cite{sec2.b1,sec2.b3,sec2.b6}. These works discuss model families, learning paradigms, and application domains spanning communications, sensing, and network management. They highlight the role of foundation models in fault diagnosis, monitoring, causal inference, and optimization, and propose high-level visions for AI-native networks. However, intelligence is largely treated as centralized inference rather than autonomous agency. Agentic reasoning, coordination mechanisms, and protocol-level integration receive limited attention. As a result, closed-loop autonomy, multi-agent orchestration, and lifecycle management remain insufficiently addressed.

A second cluster of studies focuses on conceptual and theoretical foundations of Agentic AI, including formalizations of autonomy, memory, goal-driven reasoning, and agent taxonomies \cite{sec2.b4,sec2.b10,sec2.b14}. Complementary works develop evaluation frameworks and taxonomies for LLM agents, distinguishing objectives, processes, and assessment dimensions \cite{sec2.b9,sec2.b10}. While these contributions provide essential conceptual clarity and methodological structure, they are largely domain agnostic. They do not map agentic constructs to telecom network management, control planes, or orchestration layers, nor do they consider telecom-specific Key Performance Indicators (KPIs), latency constraints, reliability requirements, or regulatory obligations.

A third category of surveys focuses on security and protocol aspects of agentic systems, addressing agent communication, multi-agent coordination, emerging Model Context Protocol (MCP) and Agent-to-Agent (A2A) paradigms, and vulnerabilities such as prompt injection and trust violations \cite{sec2.b2,sec2.b15}. While these works explicitly consider inter-agent interactions, they remain largely detached from telecom architectures and operational constraints. Integration with carrier-grade control frameworks, network management, and real-time radio access processes is not examined. Related analytical studies on single-agent and multi-agent designs explore coordination trade-offs \cite{sec2.b13}, but do not embed these mechanisms within telecom stacks or end-to-end network architectures.

Visionary and forward-looking contributions propose Artificial General Intelligence (AGI) native wireless systems, cognitive telecom brains, and fully autonomous network control \cite{sec2.b5,sec2.b8, sec2.b16}. Tutorials on LLM-centric design and multi-agent frameworks for 6G attempt to bridge Agentic AI and telecom systems \cite{sec2.b7, sec2.b11}. While these works are valuable in articulating long-term trajectories and high-level design principles, they remain largely qualitative, with limited treatment of evaluation benchmarks and standardization pathways. Their speculative nature and lack of technical details reduce their utility for near-term engineering and deployment. 


\textbf{Current Research Gaps \& Motivation.} Existing surveys either focus on LLM-enabled networking without agentic depth \cite{sec2.b1, sec2.b3,sec2.b6}, or develop agentic theory and evaluation frameworks without grounding in telecom architectures and standards \cite{sec2.b4,sec2.b9,sec2.b10,sec2.b14}, or analyze security and protocols without domain-specific integration \cite{sec2.b2,sec2.b15}. Visionary works outline AGI native networks but lack actionable detail \cite{sec2.b5,sec2.b8}, while tutorials and analytical studies provide partial bridges that remain insufficiently comprehensive \cite{sec2.b11,sec2.b13}.
 \textbf{To the best of our knowledge, no prior work jointly and systematically addresses LLMs, agentic concepts, agentic protocols, evaluation methodologies, benchmarks, telecom network integration, and standardization initiatives in a unified study.} This gap motivates the need for a holistic survey that positions LLM-powered Agentic AI as a first-class architectural paradigm for next-generation 5G and 6G network management, spanning protocol design, autonomous control, performance evaluation, and alignment with ongoing industry and standardization efforts.
Our survey addresses these deficiencies by offering the first comprehensive and tutorial-style study of LLM-powered Agentic AI with explicit emphasis on next-generation telecommunications.

\subsection{Survey Contributions}

The key contributions of this survey can be  summarized as follows:

\begin{itemize}
    \item \textbf{Comprehensive technical treatment of Agentic AI for 5G/6G Networks:}
This survey adopts a tutorial-first and review-driven methodology that establishes a unified conceptual foundation before delivering a concise yet rigorous technical survey of Agentic AI in the context of 5G and 6G network management.


    \item \textbf{Cross-domain tutorial perspective:}
    The survey is structured to be accessible to both AI and telecommunications communities, providing network researchers with essential background on LLMs, agentic reasoning, and autonomy, while offering AI researchers a clear overview of networking fundamentals, thereby fostering cross-disciplinary understanding.

    \item \textbf{Review of emerging agentic protocols and evaluation benchmarks:}
    The survey analyzes emerging Agentic AI protocol frameworks, including MCP and A2A Communication, from a telecommunications perspective, and reviews relevant evaluation and benchmarking approaches for agentic networking systems.

    \item \textbf{Unified taxonomy of LLM-based Agentic AI applications in telecom:}
    A compact taxonomy is introduced to categorize Agentic AI applications across the telecom ecosystem, covering infrastructure, management, and AI-native control domains, including radio, core, transport, and edge environments, as well as ZSM, IBN, Explainable AI (XAI), and AI Operations (AIOps).

    \item \textbf{Standardization, industry relevance, and future directions:}
    The survey summarizes ongoing standardization and industry initiatives related to AI-native and autonomous networks, and identifies key challenges and research directions for the design and deployment of next-generation agentic networking systems.
\end{itemize}

\begin{figure*}[ht!]
\centering
\includegraphics[width=0.85\textwidth]{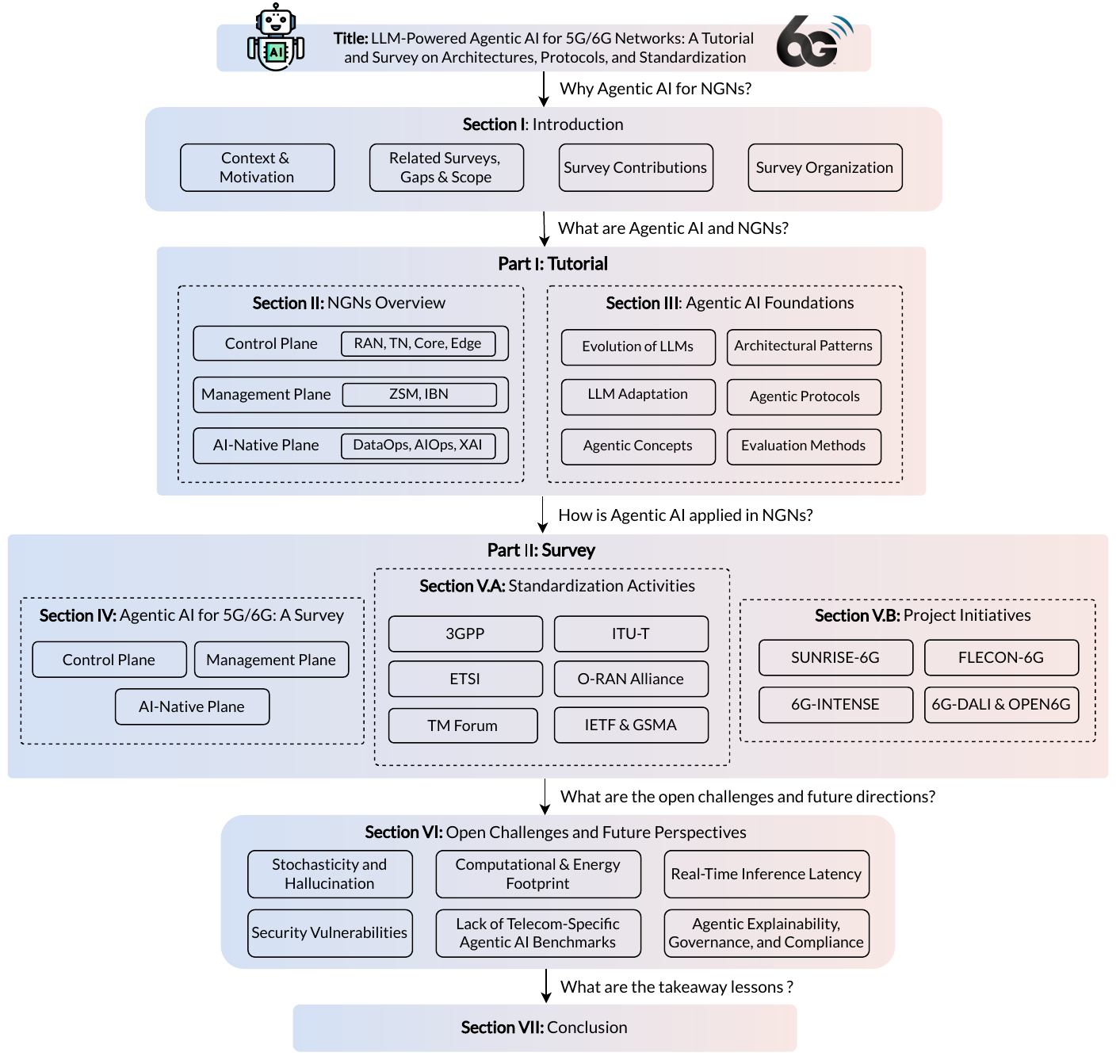}
\caption{Survey Structure.}
\label{fig-organization}
\end{figure*}

\subsection{Survey Organization}
Fig. \ref{fig-organization} depicts the structure of this paper, which combines a tutorial foundation with a state-of-the-art survey. Section \ref{sec:ngn-tuto} scopes the architectural foundations of NGNs across the radio access, transport, core, and edge/cloud domains, together with the IBN and ZSM management paradigms. Section \ref{sec:agentic-tuto} covers core concepts, architectural patterns, inter-agent communication, and evaluation methodologies relevant to networking. Section \ref{sec:agentic-for-ngn} surveys Agentic AI applications across telecommunication layers, spanning infrastructure intelligence, management and orchestration, and AI-native control. Section \ref{sec:standards-projects} reviews current standardization activities and major 6G research initiatives and their convergence toward agentic design principles. Section \ref{sec:open-challenges} discusses open challenges and future research directions, and Section \ref{sec:conclusion} concludes the survey.

\section{Next-Generation Networks: An Overview}
\label{sec:ngn-tuto}

\begin{figure}[ht!]
    \centering
    \includegraphics[width=\textwidth]{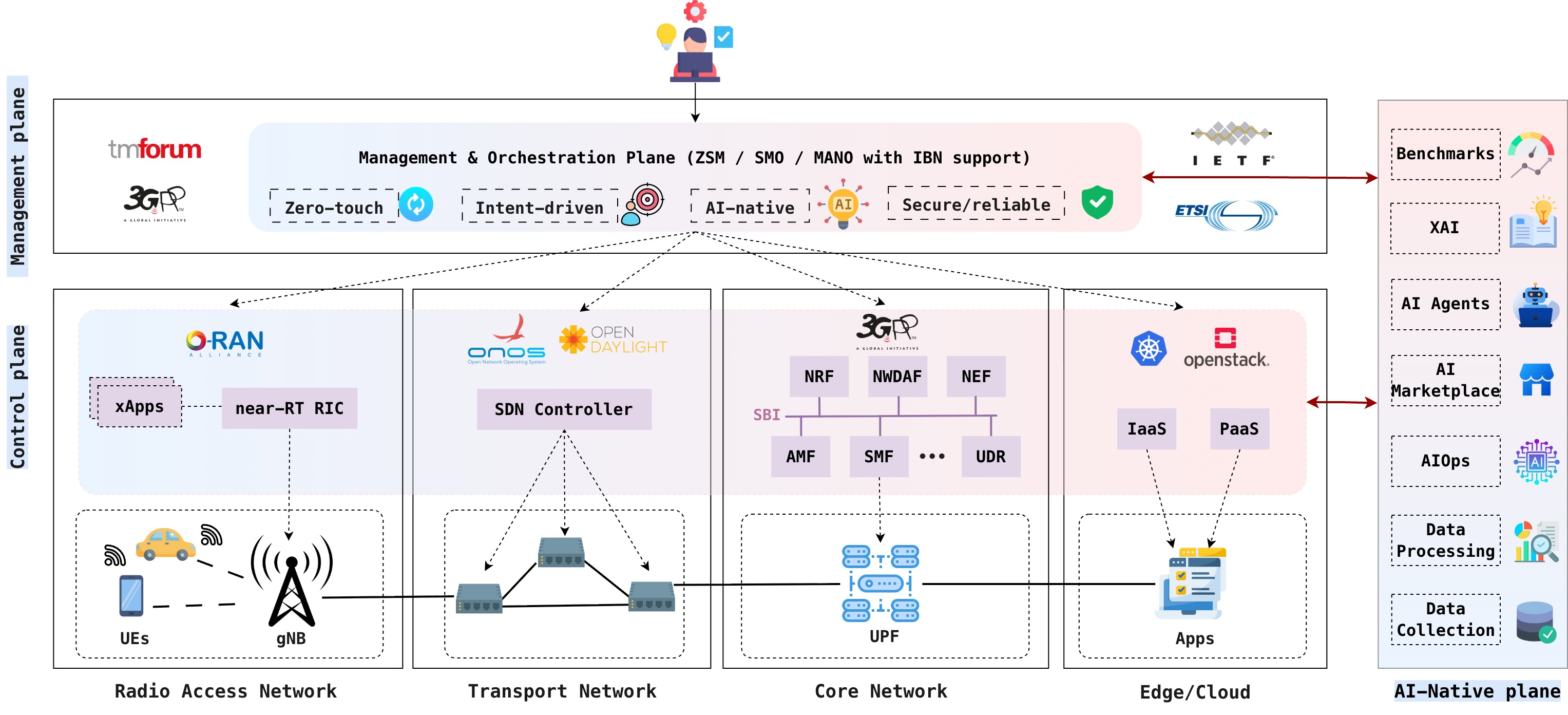}
    \caption{Overview of AI-native 6G technologies.}
    \label{fig:6g-layers}
\end{figure}

While prior generations from 1G through 5G progressively introduced digital transmission, mobile broadband, and softwarization, 6G is expected to be distinguished by the pervasive integration of intelligence across all architectural layers \cite{giordani2020toward}. As in 5G, the 6G control plane will continue to host a subset of management-related functionalities, increasingly augmented by AI-driven mechanisms for tasks such as mobility management, resource allocation, and network optimization. In parallel, the management plane, encompassing paradigms such as ZSM and IBN, provides higher-level orchestration, automation, and policy enforcement capabilities. These management functionalities inherently span both planes, reflecting their cross-layer nature. Therefore, this section focuses on the architectural components most relevant to the deployment of Agentic AI for 6G network management. As illustrated in Fig.~\ref{fig:6g-layers}, we organize the discussion around three conceptual layers: the \textit{control plane}, the \textit{management plane}, and a vertical \textit{AI-native plane} that intersects both. This abstraction aligns with the AI-native vision advocated in recent standardization efforts \cite{shehzad2022artificial}.

\subsection{Control plane}
\label{subsec:infra}

In next-generation networks, the control plane is decomposed across multiple technological domains, namely the Radio Access Network (RAN), Transport Network (TN), Core Network (CN), and Edge/Cloud infrastructure. Each of these domains adopts the principle of Control and User Plane Separation (CUPS), enabling independent scaling, modularization, and programmable control of Network Functions (NFs) \cite{choi2022ran}. In this work, we focus on the control plane, as it increasingly hosts AI-driven functionalities supporting a wide range of management tasks.

In the \textbf{RAN}, the O-RAN initiative has redefined the control plane through the RAN Intelligent Controller (RIC), which exposes two programmable surfaces: the Near-Real-Time (Near-RT) RIC for sub-second control loops via xApps, and the Non-Real-Time (Non-RT) RIC for policy-driven loops via rApps \cite{azariah2024survey}. These constructs allow third-party agents to manipulate radio resource allocation, beamforming, and handover decisions, which is essential for embedding agentic reasoning at the edge of the network. On the data plane side, 6G is expected to extend this programmability to mmWave, Terahertz bands, and cell-free massive MIMO architectures \cite{gupta2015survey}.

The \textbf{TN} adopts Software-Defined Networking (SDN) to centralize routing decisions in a controller that maintains a global topology view, enabling dynamic path computation, slicing, and load balancing \cite{long2022software}. Mainstream controllers such as ONOS\footnote{\url{https://github.com/opennetworkinglab/onos/}}, OpenDaylight\footnote{\url{https://www.opendaylight.org}}, and the 6G-oriented TeraFlowSDN\footnote{\url{https://tfs.etsi.org}} expose northbound APIs through which AI agents can inject routing intents or trigger reconfiguration based on predicted traffic patterns \cite{prabha2022survey}. Performance-critical capabilities such as Segment Routing (SRv6) and Time-Sensitive Networking (TSN) provide the deterministic substrate over which agentic decisions are enforced.

The \textbf{CN} follows a Service-Based Architecture (SBA) in which NFs communicate through standardized HTTP/2 APIs, providing a natural integration point for AI agents \cite{husain20223gpp}. Key NFs include the Access and Mobility Management Function (AMF), Session Management Function (SMF), and the User Plane Function (UPF) accelerated through eBPF/XDP for kernel-level packet processing \cite{do2021kernel}. Of particular relevance to Agentic AI are the Network Data Analytics Function (NWDAF), which supplies analytics events such as mobility prediction or abnormal traffic detection \cite{mekrache2023combining, ameur2026mlops}, and the Network Exposure Function (NEF), which provides a secure boundary through which external agents can interact with CN services \cite{fragkos2022nefsim}.

The \textbf{Edge/Cloud} domain hosts vertical applications and is governed by container orchestration platforms such as Kubernetes\footnote{\url{https://kubernetes.io}} and OpenShift\footnote{\url{https://github.com/openshift}}, which expose declarative APIs for workload placement, scaling, and lifecycle management \cite{younis2024comprehensive}. This declarative interface is well-suited for Agentic AI, since high-level placement decisions, including the dynamic migration of latency-sensitive workloads between centralized clouds and edge sites, can be expressed as desired-state manifests that the orchestrator reconciles \cite{kitanov2024overview}.

\subsection{Management plane}
\label{subsec:mgmt}

The management plane orchestrates services end-to-end across the four infrastructure domains. Given the scale and complexity of 6G, manual management is impractical, motivating two complementary paradigms that are increasingly addressed through Agentic AI and constitute the focus of this survey: IBN and ZSM. These paradigms are jointly shaped by the IETF\footnote{\url{https://www.ietf.org}}, TM Forum, 3GPP, and ETSI, and although they overlap in scope, we describe them separately for clarity.

\subsubsection{Intent-Based Networking}

IBN abstracts network governance to high-level declarative goals, called \textit{intents}, that specify the desired outcome rather than the configuration steps required to achieve it \cite{sec1.ref5}. The IBN lifecycle consists of five canonical stages: (i) \textit{profiling}, in which the requirement is captured; (ii) \textit{translation}, in which the intent is decomposed into low-level configurations targeting the RAN, TN, CN, and Edge/Cloud; (iii) \textit{resolution}, in which conflicts between competing resource requests are reconciled; (iv) \textit{activation}, in which the configurations are deployed; and (v) \textit{assurance}, in which compliance is continuously monitored. A representative intent, such as instantiating an end-to-end service composed of a gNB, a CN slice, and an edge application with 1\,Gbps guaranteed throughput and ultra-low latency, illustrates the cross-domain coordination required. Current standards encode intents in structured formats such as JSON or YAML, yet these still impose data-model expertise on operators. Next-generation IBN therefore moves toward natural-language interfaces enabled by Agentic AI, allowing operators to express intents conversationally and delegate the technical decomposition to autonomous agents \cite{mekrache2024intent}.

\subsubsection{Zero-touch Service Management}

ZSM, standardized by ETSI, targets fully autonomous network operation in which detection, prediction, and remediation occur without human intervention \cite{coronado2022zero,sec6.b4}. When combined with IBN, ZSM operationalizes the assurance stage by enforcing intent compliance throughout the service lifecycle. The closed automation loop comprises three stages: (i) \textit{anomaly detection and prediction}, typically driven by time-series AI models that flag degradations before they impact users; (ii) \textit{Root Cause Analysis (RCA)}, increasingly performed with explainable AI to expose the causal chain behind a fault \cite{dwivedi2023explainable}; and (iii) \textit{anomaly resolution}, which requires planning and reasoning to select corrective actions. The reasoning requirement of the third stage explains the recent surge of interest in Agentic AI for ZSM, since LLM-based agents can synthesize multi-step remediation plans grounded in current network state.

\subsection{AI-native plane}
\label{subsec:ai-layer}

The AI-native plane is a vertical pillar that supplies intelligence to both the infrastructure and the management layers. In the control plane, it manifests as xApps and rApps in the RIC for radio control, AI-augmented SDN routing, NWDAF-driven analytics in the CN, and predictive workload placement in the Edge/Cloud. In the management plane, it underpins natural-language intent translation, conflict resolution, and the closed-loop assurance enabled by ZSM. Standards bodies including TM Forum and ETSI have articulated this AI-native vision while emphasizing that explainability is a prerequisite for autonomous decision-making \cite{wu2021toward}. We organize the AI-native layer around three pillars: \textit{Data Operations}, \textit{AI Operations (AIOps)}, and \textit{Explainability}.

\textbf{Data Operations.} High-quality data is the prerequisite of any AI-driven function, and the AI-native layer must collect, store, and preprocess telemetry from across all domains \cite{chochliouros2025developing}. Domain-specific telemetry pipelines include monitoring xApps in the RAN \cite{santos2025managing}, In-band Network Telemetry (INT) in the TN \cite{tan2021band}, NWDAF events and 3GPP Event Exposure APIs in the CN \cite{mekrache2023combining}, and Prometheus-based metrics in the Edge/Cloud \cite{ksentini2025lightweight}. Beyond raw telemetry, human feedback is becoming a first-class data source for IBN, enabling continuous retraining of intent translation models to improve user satisfaction.

\textbf{AI Operations.} AIOps cover the lifecycle of AI models, from offline training on historical datasets to online adaptation under dynamic conditions \cite{kuklinski2025mlops}. Federated Learning is increasingly adopted to train models locally at the edge while preserving data privacy and capturing global patterns \cite{yang2022federated}. On the inference side, the millisecond budget of 6G control loops elevates the importance of optimization techniques such as pruning, quantization, and Knowledge Distillation, the last allowing a compact student model to approximate a larger teacher with minimal accuracy loss \cite{ref6.5,fang2026knowledge}.

\textbf{Explainability.} Black-box AI is incompatible with the accountability requirements of operational networks and with regulatory frameworks such as the EU AI Act. Explainable AI techniques therefore play a central role in the AI-native layer: SHAP \cite{antwarg2021explaining} and LIME \cite{zafar2021deterministic} attribute decisions to specific input features such as traffic load or latency, while Agentic AI itself is increasingly leveraged to produce natural-language explanations of complex network states \cite{mekrache2024combining}, closing the loop between automation and human oversight.

\section{Agentic AI Foundations}
\label{sec:agentic-tuto}
This section reviews and analyzes recent advancements in LLM-based Agentic AI from a telecom perspective, with emphasis on their implications for future 5G/6G Networks. 


\subsection{From LLMs to Agentic AI}

\begin{figure}[t]
    \centering
    \includegraphics[width=0.96\textwidth]{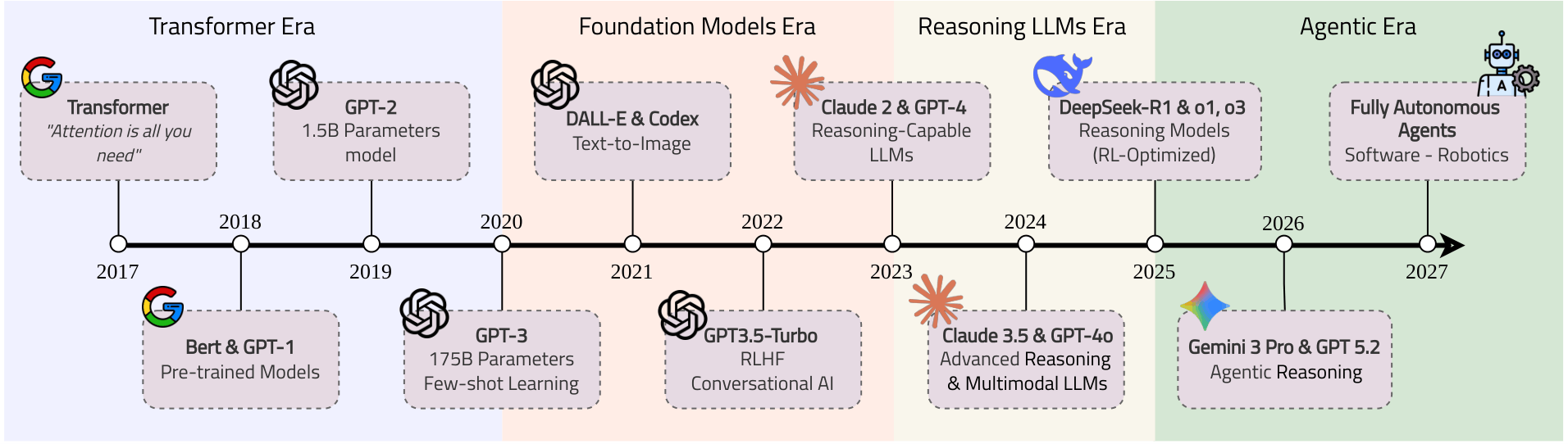}
    \caption{LLM-based Agentic AI paradigm Evolution.}
    \label{fig:agnetic-ai-evolution}
\end{figure}

The progression from LLMs to Agentic AI represents a shift from static, prompt-driven predictors to systems capable of autonomous operation within complex and evolving environments \cite{sec2.b6, sec2.b15}.
As shown in Fig. \ref{fig:agnetic-ai-evolution}, this shift follows a clear architectural trajectory that begins with transformer-based and foundation models, progresses through reasoning-optimized LLMs, and culminates in agentic systems that integrate language understanding with decision-making and control.
While modern LLMs exhibit strong generalization, reasoning, and multimodal understanding, these abilities alone are insufficient for domains that require sustained interaction, adaptive decision processes, and operational accountability \cite{sec2.b8}. NGNs exemplify such domains, where tasks involve continuous monitoring, intent-driven optimization, coordinated control across layers, and robust handling of uncertainty \cite{sec2.b7}. Agentic AI emerges as an architectural response that integrates LLMs with memory, planning, tool access, and alignment mechanisms to support reliable and context aware autonomy \cite{sec2.b7,sec2.b10}.

\subsection{LLM Adaptation Strategies for Agentic Workflows}

Adaptation strategies transform general-purpose LLMs into reliable, context-aware components
for intelligent network management. Formally, let $f_{\theta}: \mathcal{X}
\rightarrow \mathcal{Y}$ denote a pre-trained LLM with parameters $\theta \in \mathbb{R}^{d}$.
Adaptation seeks a task-specialized predictor $f_{\theta'}(\cdot\,;\,c)$, where $\theta'$
denotes (possibly updated) parameters and $c$ an in-context conditioning signal. Two
complementary paradigms are employed: \textbf{Fine-Tuning (FT)}, which modifies
$\theta \rightarrow \theta'$ via optimization on task data, and \textbf{In-Context Learning
(ICL)}~\cite{sec3.b13}, which fixes $\theta' = \theta$ and adapts behavior through $c$ alone.

Table~\ref{tab:llm_adaptation} summarizes both paradigms, their core mechanisms, and
telecom alignment.

\begin{table*}[ht]
\centering
\caption{Taxonomy of LLM Adaptation Strategies and their Telecom Alignment.}
\label{tab:llm_adaptation}
\small
\renewcommand{\arraystretch}{1.05}
\begin{tabular}{p{1.6cm} p{3.6cm} p{4.4cm} p{4.6cm}}
\hline
\rowcolor{lightgray!25}
\textbf{Paradigm} & \textbf{Technique} & \textbf{Core Mechanism} & \textbf{Telecom Alignment} \\
\hline\hline
\multirow{5}{*}{\textbf{Fine-Tuning}} 
  & SFT \cite{sec3.b14}
  & Cross-entropy minimization on labeled pairs $\mathcal{D}$
  & Fault diagnosis, traffic classification, KPI prediction \\
\cmidrule(lr){2-4}
  & PEFT / LoRA \cite{sec3.b15,ref55}
  & Low-rank update $\Delta W{=}BA$; optimizes $\phi$, $|\phi|{\ll}d$
  & Efficient adaptation at edge/RAN with limited compute \\
\cmidrule(lr){2-4}
  & RL / RLHF / GRPO \cite{sec3.b16}
  & KL-regularized reward maximization: $\mathbb{E}[r_\psi] - \beta\,\mathrm{KL}$
  & QoS/latency alignment; policy-compliant decision-making \\
\cmidrule(lr){2-4}
  & Knowledge Distillation \cite{sec3.b17,ref6.5}
  & $\min\,\mathrm{KL}(p_{\theta_T}\|p_{\theta_S})$ from teacher to student
  & Compact models for base stations and edge inference \\
\cmidrule(lr){2-4}
  & ReFT \cite{sec3.b18}
  & Intervention $\Phi_\phi$ on hidden states $h^{(\ell)}$; $|\phi|{\ll}d$
  & Traffic dynamics and anomaly signature modeling \\
\midrule
\multirow{5}{*}{\parbox{1.6cm}{\textbf{In-Context\\Learning}}}
  & Zero-Shot \cite{sec3.b19}
  & $c {=} \mathcal{I}$; no examples or retraining
  & Intent-based management; NL network configuration queries \\
\cmidrule(lr){2-4}
  & Few-Shot \cite{sec3.b20}
  & $c {=} \{\mathcal{I},(x_j,y_j)_{j=1}^k\}$; small-$k$ exemplars
  & Alarm/log-guided classification and summarization \\
\cmidrule(lr){2-4}
  & CoT / ToT \cite{sec3.b21}
  & Marginalizes over latent steps $z{=}(z_1,\ldots,z_T)$
  & Root cause analysis; multi-step troubleshooting workflows \\
\cmidrule(lr){2-4}
  & RAG \cite{sec3.b22}
  & Top-$k$ retrieval $\mathcal{R}(x)$ from corpus $\mathcal{K}$
  & Standards/spec grounding; up-to-date operational knowledge \\
\cmidrule(lr){2-4}
  & Tool / API Integration \cite{sec3.b23}
  & Action $a{=}(\tau,u)$; generation conditioned on $o{=}\tau(u)$
  & OSS/BSS and SDN closed-loop control; real-time monitoring \\
\bottomrule
\end{tabular}
\end{table*}

\textbf{Fine-Tuning} methods~\cite{sec3.b12} optimize over a labeled telecom corpus
$\mathcal{D} = \{(x_i, y_i)\}$. Supervised Fine-Tuning (SFT)~\cite{sec3.b14} minimizes
token-level cross-entropy: $\theta^{\ast} = \arg\min_{\theta} \frac{1}{N}\sum_{i}
\mathcal{L}(f_{\theta}(x_i), y_i)$, specializing the model to domain tasks such as fault
diagnosis and traffic classification. Parameter-Efficient Fine-Tuning
(PEFT)~\cite{sec3.b15}, including Low-Rank Adaptation (LoRA)~\cite{ref55}, which
decomposes weight updates as $\Delta W = BA$ with rank $r \ll \min(m,n)$ that restricts
optimization to a low-dimensional subspace $\phi \in \mathbb{R}^{k}$, $k \ll d$, enabling
deployment at edge and RAN nodes with limited compute. Reinforcement and reward-based
methods~\cite{sec3.b16} (e.g., Reinforcement Learning Human Feedback (RLHF), Group Relative Policy Optimization (GRPO)) treat the LLM as a stochastic policy
$\pi_{\theta}(y \mid x)$ and maximize expected reward under a KL-regularized objective
$\mathbb{E}[r_{\psi}(x,y)] - \beta\,\mathrm{KL}(\pi_{\theta}\|\pi_{\mathrm{ref}})$, aligning
outputs with QoS and operational constraints. Knowledge Distillation~\cite{sec3.b17,ref6.5}
minimizes $\mathrm{KL}(p_{\theta_T}\|p_{\theta_S})$ to transfer capabilities from a large
teacher to a compact student suitable for base-station or edge deployment. Representation
Fine-Tuning (ReFT)~\cite{sec3.b18} instead learns an intervention
$\Phi_{\phi}: h^{(\ell)} \mapsto \tilde{h}^{(\ell)}$ on hidden states, capturing
telecom-specific patterns (e.g., traffic dynamics, anomaly signatures) without full
weight updates.

\textbf{In-Context Learning} fixes $\theta$ and adapts $f$ through the conditioning context
$c$, yielding $\hat{y} = \arg\max_{y} p_{\theta}(y \mid c, x)$~\cite{sec3.b13}. Zero-shot
prompting~\cite{sec3.b19} sets $c = \mathcal{I}$ (a task instruction), enabling
intent-driven network queries without retraining. Few-shot prompting~\cite{sec3.b20}
augments $c$ with $k$ telecom examples $\{(x_j, y_j)\}_{j=1}^{k}$, improving performance
on classification, summarization, and diagnosis. Chain-/Tree-of-Thought
(CoT/ToT)~\cite{sec3.b21} introduces latent reasoning steps $z = (z_1,\ldots,z_T)$,
marginalizing $p_{\theta}(y \mid c, x) = \sum_{z} p_{\theta}(y \mid z,c,x)\,p_{\theta}(z
\mid c,x)$ over sequential or branching trajectories for structured telecom workflows (e.g.,
root cause analysis). Retrieval-Augmented Generation
(RAG)~\cite{sec3.b22} conditions generation on top-$k$ documents retrieved from a telecom
knowledge corpus $\mathcal{K}$ including standards, specifications, operational logs, via a retriever
$\mathcal{R}: \mathcal{X} \rightarrow \mathcal{K}^{k}$, ensuring factual grounding without
parameter modification. Finally, Tool/API Integration~\cite{sec3.b23} exposes a set
$\mathcal{T} = \{\tau_1,\ldots,\tau_M\}$ of telecom system interfaces (e.g. Operations Support System (OSS) and Business Support System (BSS), SDN
controllers), letting the model select action $a = (\tau, u)$ and condition subsequent
generation on the observation $o = \tau(u)$, enabling closed-loop automation.

\subsection{Agentic AI Core Concepts}

\begin{figure}[t]
    \centering
    \includegraphics[width=0.75\textwidth]{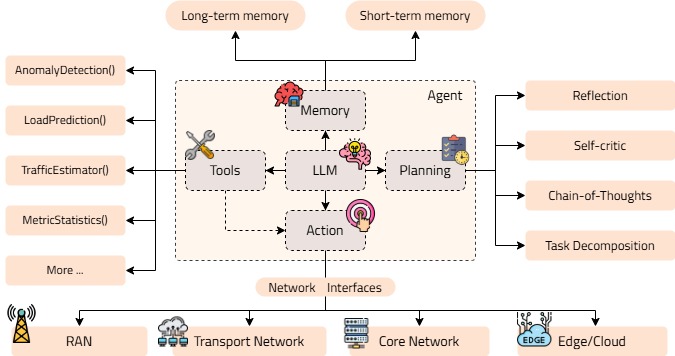}
    \caption{Telecom-oriented Agentic AI Framework Overview.}
    \label{fig:agnetic-ai-framework}
\end{figure}

Agentic AI introduces a set of architectural abstractions that extend LLMs from passive inference engines to autonomous decision-making systems \cite{sec2.b2}.
As illustrated in Fig. \ref{fig:agnetic-ai-framework}, the core components include the agent powered by LLMs, memory systems, planning mechanisms, tool interfaces, and action execution pathways \cite{sec2.b14}. Short-term and long-term memory support contextual grounding and continuity across tasks. Planning modules provide structured reasoning capabilities, including reflection, self-critiquing, and task decomposition. Tool interfaces enable the agent to invoke analytical and operational functions such as anomaly detection, load prediction, and traffic estimation. The resulting actions are executed through standardized interfaces spanning the CN, RAN, and cloud/edge domains \cite{sec2.b4, sec2.b11}. Collectively, these components form a modular and interoperable foundation that governs the autonomy, adaptability, and reliability of agentic systems operating in NGNs.

Formally, an agentic system can be represented as a tuple:
\begin{equation}
\mathcal{A} = \langle \mathcal{S}, \mathcal{O}, \mathcal{A}_c, \mathcal{M}, \mathcal{T}, \pi_{\theta}, \mathcal{G} \rangle,
\end{equation}
where $\mathcal{S}$ is the environment state space (e.g., CN/RAN/edge conditions), $\mathcal{O}$ the observation space, $\mathcal{A}_c$ the action space, $\mathcal{M}$ the memory system, $\mathcal{T}$ the set of available tools, $\pi_{\theta}$ the LLM-based policy, and $\mathcal{G}$ the goal specification. The following subsections delve into the details of these concepts.

\subsubsection{Agents}
An agent is an autonomous computational system that interprets observations, maintains state, makes decisions, and performs actions to achieve defined goals \cite{sec2.b2, sec3.b26}. In LLM-based settings, agents extend pretrained models with planning, tool use, and persistent memory, enabling them to operate through iterative control loops rather than single-step inference. This differentiates agentic systems from conventional conversational LLMs, which lack self-directed action and sustained contextual reasoning \cite{sec2.b4, sec2.b11}.

Formally, at each decision step $t$, the agent receives an observation $o_t \in \mathcal{O}$, updates its internal memory state $m_t \in \mathcal{M}$, and selects an action $a_t \in \mathcal{A}_c$ through an LLM-parameterized policy:
\begin{equation}
a_t \sim \pi_{\theta}\!\left(a_t \,\big|\, o_t,\, m_t,\, \mathcal{G}\right), \qquad m_{t+1} = \mathcal{U}(m_t, o_t, a_t),
\end{equation}
where $\mathcal{U}$ denotes the memory update operator. The agent seeks to maximize a cumulative goal-aligned utility
\begin{equation}
J(\pi_{\theta}) = \mathbb{E}_{\pi_{\theta}}\!\left[\sum_{t=0}^{H} \gamma^{t}\, \mathcal{R}(s_t, a_t \,;\, \mathcal{G})\right],
\end{equation}
with horizon $H$, discount factor $\gamma \in (0,1]$, and goal-conditioned reward $\mathcal{R}$ capturing objectives such as QoS satisfaction or SLA compliance. Leading industry bodies like OpenAI\footnote{https://openai.com/fr-FR/} and Anthropic\footnote{https://www.anthropic.com/} emphasize that agents can dynamically structure their workflows, select tools, and adapt strategies in response to changing conditions \cite{sec3.b26, sec3.b27}. Agent architectures span reactive, deliberative, and hybrid forms, allowing a balance between rapid response and long-horizon reasoning \cite{sec2.b4}. 


\subsubsection{Memory}

Memory is a foundational component of agentic intelligence, enabling temporal continuity, contextual reasoning, and adaptive decision-making across dynamic telecom environments. Native LLMs are constrained by fixed context windows and stateless inference, whereas agentic systems introduce hierarchical memory layers to persist and structure information across time and control loops. Formally, the memory system is decomposed as \cite{sec2.b4, sec3.b26}:
\begin{equation}
\mathcal{M} = \mathcal{M}_{\text{ST}} \cup \mathcal{M}_{\text{LT}}, \quad \mathcal{M}_{\text{LT}} = \mathcal{M}_{\text{ep}} \cup \mathcal{M}_{\text{sem}} \cup \mathcal{M}_{\text{proc}},
\end{equation}
where $\mathcal{M}_{\text{ST}}$ and $\mathcal{M}_{\text{LT}}$ denote short- and long-term memory, and $\mathcal{M}_{\text{ep}}, \mathcal{M}_{\text{sem}}, \mathcal{M}_{\text{proc}}$ denote episodic, semantic, and procedural components.

\textbf{Short-term memory (STM) }corresponds to transient, high-resolution context maintained within or alongside the prompt window. It evolves as a bounded sliding buffer:
\begin{equation}
\mathcal{M}_{\text{ST}}^{(t)} = \{(o_{\tau}, a_{\tau}, r_{\tau})\}_{\tau = t-W+1}^{t}, \quad |\mathcal{M}_{\text{ST}}^{(t)}| \leq W,
\end{equation}
where $W$ is the effective context capacity. In 5G/6G management, STM captures real-time telemetry streams (e.g., KPIs, alarms, channel conditions, traffic load) and intermediate reasoning states during tasks such as fault localization or resource allocation. It enables fine-grained correlation across recent events, supports multi-step reasoning (e.g., root cause analysis pipelines), and facilitates rapid reaction in near-real-time control loops (e.g., RAN scheduling, slicing adaptation). However, STM is inherently limited in capacity and temporal scope, requiring efficient summarization and filtering mechanisms \cite{sec3.b26, sec3.b27}.

\textbf{Long-term memory (LTM)} provides persistent storage of structured and unstructured knowledge across sessions and time horizons \cite{sec3.b26, sec3.b27}. It is typically realized as an indexed key–value store $\mathcal{M}_{\text{LT}} = \{(k_i, v_i)\}_{i=1}^{N}$, where $k_i \in \mathbb{R}^{d_e}$ is an embedding key and $v_i$ the stored content. Retrieval given a query $q$ is performed via a similarity-based selection \cite{sec2.b4}:  
\begin{equation}
\mathcal{R}(q; \mathcal{M}_{\text{LT}}) = \underset{(k_i, v_i) \in \mathcal{M}_{\text{LT}}}{\text{Top-}k}\; \mathrm{sim}\!\left(\phi(q),\, k_i\right),
\end{equation}
with $\phi(\cdot)$ an embedding function and $\mathrm{sim}(\cdot,\cdot)$ a similarity metric (e.g., cosine). In telecom settings, LTM includes historical network states, configuration changes, policy rules, learned traffic patterns, and prior optimization outcomes. It is typically implemented via external vector databases, knowledge graphs, or time-series repositories, enabling retrieval-augmented reasoning and continual learning. LTM supports trend analysis, seasonal pattern recognition, and predictive modeling (e.g., congestion forecasting, anomaly anticipation), which are critical for proactive and closed-loop network management in 6G systems.

Specialized memory forms further enhance functionality. Episodic memory $\mathcal{M}_{\text{ep}} = \{e_j = (s_j, a_j, r_j, s_j')\}$ stores past operational events (e.g., incident-response traces), enabling experience replay and case-based reasoning. Semantic memory $\mathcal{M}_{\text{sem}}$ encodes domain knowledge such as standards, protocols, and network models (often as a knowledge graph $\mathcal{G}_{\text{sem}} = (\mathcal{V}, \mathcal{E})$), supporting consistent and interpretable decisions. Procedural memory $\mathcal{M}_{\text{proc}} = \{\pi_j : \mathcal{S} \rightarrow \mathcal{A}_c\}$ captures learned action policies and workflows (e.g., automated remediation sequences), enabling execution of complex control strategies.
The integration of STM and LTM is particularly critical in telecom agentic systems, where decisions must balance immediate network conditions with long-term objectives such as QoS guarantees, energy efficiency, and SLA compliance \cite{sec3.b26, sec3.b27}.



\subsubsection{Tools}
Tool use is a defining mechanism that expands the operational scope of Agentic AI by enabling interaction with external systems beyond language generation. Through interfaces to Application Programming Interfaces (APIs), controllers, simulators, and analytical engines, agents can retrieve measurements, perform computations, and execute control actions \cite{sec3.b28}. Formally, let $\mathcal{T} = \{\tau_1, \ldots, \tau_M\}$ be the set of available tools, where each tool $\tau_i: \mathcal{U}_i \rightarrow \mathcal{O}_i$ maps an argument space $\mathcal{U}_i$ to an observation space $\mathcal{O}_i$. At step $t$, the agent jointly selects a tool and its invocation arguments \cite{sec2.b4}:
\begin{equation}
(\tau_t, u_t) \sim \pi_{\theta}\!\left(\cdot \,\big|\, o_t, m_t, \mathcal{G}\right), \quad o_t^{\text{tool}} = \tau_t(u_t),
\end{equation}
and integrates the resulting observation $o_t^{\text{tool}} \in \mathcal{O}_{\tau_t}$ into subsequent reasoning. Effective tool use requires accurate selection, invocation, and interpretation of outputs, providing grounding between internal reasoning and external environments \cite{sec3.b27}. In communication networks, tool-enabled agents support telemetry analysis, diagnostics, configuration management, and policy evaluation, allowing agents to translate high-level intents into actionable network operations.

\subsubsection{Planning}
Planning is a core capability of Agentic AI that enables the translation of high-level intents into structured and executable action sequences. Given a goal $\mathcal{G}$ and current context $c_t$, the planner produces a plan:
\begin{equation}
P = (p_1, p_2, \ldots, p_K) \sim \pi_{\theta}^{\text{plan}}(P \mid \mathcal{G}, c_t),
\end{equation}
where each subtask $p_k = (\text{op}_k, \text{pre}_k, \text{post}_k)$ specifies an operation together with its preconditions and postconditions, subject to a partial order $\prec$ encoding dependencies ($p_i \prec p_j$ if $p_j$ depends on $p_i$). It decomposes complex network objectives into ordered tasks while accounting for dependencies, constraints, and cross-domain interactions. Planning operates in close coordination with memory and tool interfaces to adapt to real-time network conditions and reuse prior knowledge. Iterative refinement through feedback and self-critique supports closed-loop control and continuous adjustment, formalized as:    
\begin{equation}
P^{(i+1)} = \pi_{\theta}^{\text{plan}}\!\left(P^{(i)},\, \mathcal{F}(P^{(i)}, o_t^{\text{exec}})\right),
\end{equation}
where $\mathcal{F}$ is a feedback/critique operator evaluating execution outcomes $o_t^{\text{exec}}$, and the iteration terminates when the goal predicate $\mathcal{G}(c_t)$ is satisfied or a budget is exhausted \cite{sec2.b14, sec3.b26}. These capabilities make planning a foundational element for intent-driven and autonomous operation in AI-native 5G and 6G networks.

\subsection{Agentic AI Architectural Patterns}

Agentic AI architectural patterns define modular design abstractions for orchestrating reasoning, planning, memory, and tool interaction within LLM-based systems. These patterns formalize how agents interact with external environments and internal cognitive modules, enabling structured autonomy, improved controllability, and scalable deployment in NGNs \cite{sec3.b29}. They further provide a principled basis for balancing latency, reliability, and adaptability under telecom-grade operational constraints. The key trade-offs across different agentic AI architectural patterns in telecom scenarios are summarized in Table~\ref{tab:agentic_tradeoffs}. The following patterns represent widely adopted agentic designs in complex networked systems.

\begin{table*}[ht]
\centering
\caption{Trade-offs of Agentic AI Architectural Patterns in Telecom Scenarios.}
\label{tab:agentic_tradeoffs}
\small
\renewcommand{\arraystretch}{1.0}
\setlength{\tabcolsep}{4pt}
\begin{tabular}{p{1.5cm} p{4.5cm} p{4.0cm} p{4.5cm}}
\hline
\rowcolor{lightgray!25}
\textbf{Arch.} & \textbf{Strengths} & \textbf{Limitations} & \textbf{Best-Fit Telecom Scenarios} \\
\hline\hline

ReAct &
Low-latency iterative reasoning with tool grounding; strong for telemetry-driven decision loops. &
Limited long-horizon planning; sensitive to noisy observations. &
Real-time fault detection, KPI-driven monitoring, RAN anomaly diagnosis. \\
\hline

\rowcolor{lightgray!25}
AutoGPT &
Strong long-horizon autonomy; supports hierarchical task decomposition and self-refinement. &
High computational overhead; potential error accumulation without supervision. &
Intent-based orchestration, automated troubleshooting, lifecycle management. \\
\hline

Agentic RAG &
High factual grounding via adaptive retrieval; improves compliance and knowledge freshness. &
Retrieval latency; dependency on external knowledge quality. &
Standards compliance checks, policy validation, root cause analysis. \\
\hline

\rowcolor{lightgray!25}
Multi-Agent Systems &
High scalability and parallelism; robust to domain decomposition and partial failures. &
Coordination complexity; communication overhead; non-trivial convergence. &
Network slicing orchestration, cross-domain optimization, large-scale SON control. \\
\hline

\end{tabular}
\end{table*}

\subsubsection{ReAct (Reason + Act)}
ReAct interleaves explicit reasoning traces with tool execution, enabling stepwise grounding of decisions in external observations \cite{sec3.b30}. This tight coupling between inference and action supports iterative telemetry acquisition and corrective control, making it suitable for closed-loop network monitoring and online diagnosis in dynamic RAN and core environments.

\subsubsection{AutoGPT}
AutoGPT implements a fully autonomous loop combining task decomposition, persistent memory, and iterative self-refinement \cite{sec3.b31}. It supports long-horizon execution where objectives are recursively decomposed into subtasks, executed via tools, and updated based on feedback signals. This makes it relevant for autonomous operations such as intent-driven orchestration, multi-stage troubleshooting, and optimization workflows across distributed telecom domains.

\subsubsection{Agentic RAG}
Agentic RAG extends retrieval-augmented generation by embedding retrieval decisions within the agent’s policy loop. The system dynamically determines retrieval timing, query formulation, and source ranking, followed by verification-aware integration of external knowledge \cite{maz2025eurocomply}. This improves factual grounding and contextual precision in telecom scenarios involving standards compliance, fault localization, and performance auditing.

\subsubsection{Multi-Agent Systems}
Multi-Agent Systems (MAS) distribute reasoning and action across specialized agents operating under coordination protocols. Each agent may focus on a functional plane (e.g., radio optimization, transport control, service assurance), enabling parallelism, fault isolation, and scalable decision-making \cite{sec3.b32}. This paradigm is particularly effective for large-scale NGNs requiring cross-domain orchestration, resilience under partial failure, and concurrent optimization of heterogeneous network slices.

\subsection{Agentic AI Protocols}

The rapid proliferation of Agentic AI systems has exposed challenges in interoperability, scalability, and standardization, as early solutions relied on proprietary frameworks and ad-hoc integrations. This fragmentation hinders reproducibility and cross-system coordination, particularly in large-scale domains such as telecommunications. In response, emerging agentic protocols standardize context access, tool invocation, intent exchange, and multi-agent collaboration, forming a foundation for consistent interaction and semantic alignment across distributed systems \cite{sec2.b2,sec3.b35}.

\begin{table*}[t]
\centering
\caption{Summary of Agentic AI protocols and their relevance to Future 5G/6G Networks.}
\label{tab:agentic_ai_protocols}
\small
\renewcommand{\arraystretch}{1.0}
\setlength{\tabcolsep}{4pt}
\begin{tabular}{p{1.2cm} p{2.9cm} p{3.2cm} p{2.5cm} p{1.5cm} p{2.5cm}}
\hline
\rowcolor{lightgray!25}
\textbf{Protocol} & \textbf{Primary Objective} & \textbf{Key Mechanisms} & \textbf{6G Integration Point} & \textbf{Transport} & \textbf{Constraint} \\
\hline \hline

MCP &
Standardize LLM-based agent access to external tools and data sources. &
JSON-RPC 2.0 over stdio or Streamable HTTP/SSE; \texttt{tools/call} with tool \texttt{name} and typed \texttt{arguments}. &
Non-RT RIC, SMO, NWDAF, OSS/BSS interfaces. &
stdio (local), HTTP/1.1 + (remote, TLS) &
50--200\,ms per tool chain on loaded edge. \\
\hline

\rowcolor{lightgray!15}
A2A &
Interoperable coordination among heterogeneous LLM-based agents. &
Capability cards, task lifecycle objects, and streaming; JSON-RPC or gRPC transport. &
Non-RT and near-RT RIC; RAN optimization, slicing, anomaly detection. &
HTTP+SSE, gRPC (TLS) &
gRPC serialization overhead in high-frequency small-message exchanges. \\
\hline
ANP &
Open-internet agent discovery and decentralized collaboration. &
Decentralized identity, structured capability descriptions, web-standard discovery endpoints. &
Edge and core orchestration; hierarchical and flat agent topologies. &
HTTPS, secured P2P &
Identity resolution latency; stale routing during rapid topology changes. \\
\hline
\rowcolor{lightgray!15}
ACP &
Interoperable messaging between LLM agents and legacy network interfaces. &
Multipart typed payloads over REST; synchronous and asynchronous patterns with session management. &
Full RAN--core--edge stack; legacy NF and OSS/BSS interoperability. &
HTTP (TLS) &
Polling overhead for async flows; no native push without streaming extension. \\
\hline
AP2 &
Secure verifiable agent-initiated transactions. &
Cryptographically signed mandate chain (intent, cart, payment) as verifiable credentials. &
Business and service layers; multi-operator leasing and spectrum trading. &
HTTPS (TLS) &
Signing latency (1--2\,ms/mandate) confines use to non-RT economic loops. \\
\hline

\end{tabular}
\end{table*}

\subsubsection{Model Context Protocol (MCP)}
MCP was introduced by Anthropic in November 2024 to standardize how LLM-based applications communicate with external tools, data sources, and services through 
a formal client-server architecture~\cite{sec3.b35}. At the wire level, MCP messages are serialized as JavaScript Object Notation/Remote Procedure Call (JSON-RPC) 2.0 objects. The protocol defines two 
standard transport mechanisms: \textit{stdio}, where JSON-RPC messages are exchanged over standard input/output streams, and \textit{Streamable HTTP}, where 
client-to-server messages are sent as HTTP POST requests and server-to-client streaming is delivered via Server-Sent Events (SSE). Tool invocation is performed via a \texttt{tools/call} request, whose \texttt{params} object carries a \texttt{name} field identifying the target tool and an \texttt{arguments} field containing the typed key-value input map validated against the tool's 
\texttt{inputSchema}.
In future 5G/6G deployments, MCP has been adopted to interface LLM agents with NWDAF analytics endpoints, O-RAN xApp registries, and network configuration databases~\cite{li2025netmcp,ameur2026agentic}, primarily at the non-RT RIC and management plane, where its latency profile is compatible with control timescales.

\subsubsection{Agent-to-Agent Communication (A2A)}
Announced by Google in April 2025, A2A defines structured coordination
among heterogeneous LLM-based agents~\cite{sec3.b35}. Messages are organized
around four core objects: an Agent Card advertising capabilities, a Task with
a defined lifecycle, an Artifact carrying output, and a Message as the atomic
communication unit. Transport bindings include JSON-RPC 2.0 over HTTP with SSE
streaming, gRPC (Protocol Buffers v3), and HTTP/REST. In 6G contexts, A2A
supports decentralized collaboration for RAN optimization, slicing, and anomaly
detection, with gRPC offering lower serialization overhead than JSON-RPC for
high-frequency inter-agent exchanges~\cite{sec2.b2}.

\subsubsection{Agent Network Protocol (ANP)}
ANP models agents as networked entities that advertise capabilities and discover
peers on the open internet~\cite{sec3.b33}. Built on  Decentralized Identifiers (DIDs), Verifiable
and Credentials, it defines a three-layer architecture covering
identity, meta-protocol negotiation, and application interactions. Agent
discovery relies on the well-known URI path, returning
a JSON document listing public agent descriptions under a domain. For 5G/6G networks,
ANP supports deployment across RAN, core, and edge domains; however,
decentralized identity resolution introduces latency that can cause stale
routing during rapid topology changes, such as handovers, a failure mode requiring bounded-latency extensions not yet addressed in the literature.

\subsubsection{Agent Communication Protocol (ACP)}
Developed by IBM Research and released in May 2025 under Linux Foundation
governance, ACP defines a RESTful HTTP interface for LLM-driven agent
interoperability~\cite{sec3.b34}. Messages are Media types multipart payloads
over HTTP, supporting synchronous and asynchronous patterns with structured
session management and authentication via Role-Based Access Control (RBAC). In NGNs, ACP
facilitates intent exchange between LLM agents and legacy NF management
interfaces; its framework-agnostic REST transport and offline agent packaging
suit heterogeneous RAN--core--edge deployments without requiring persistent
connections.

\subsubsection{Agent Payments Protocol (AP2)}
Announced by Google in September 2025, AP2 standardizes secure financial
interactions among autonomous agents via cryptographically signed mandates
structured as Verifiable Credentials over HTTPS with Transport Layer Security (TLS)
~\cite{sec3.b36}. Three mandate types, Intent, Cart, and Payment, are signed
using encryption protocols, forming a non-repudiable audit chain from user
delegation to settlement. From a telecom perspective, AP2 supports multi-operator resource leasing,
spectrum trading, and autonomous service procurement~\cite{barros2025ai},
though its cryptographic overhead confines it to non-RT business-layer
transactions rather than sub-second RAN-tier resource negotiation.

\subsubsection{Integration of Agentic Protocols in 6G Networks}
Table~\ref{tab:agentic_ai_protocols} maps each protocol to its target 6G integration point. MCP aligns with the non-RT RIC and management plane, where its tool-invocation model interfaces directly with NWDAF, OSS/BSS, and O-RAN
SMO. A2A operates across the non-RT and near-RT RIC tiers, enabling inter-agent coordination for RAN optimization and slicing via JSON-RPC or gRPC transport. ANP targets edge and core orchestration, where web-standard discovery and decentralized identity support large agent collectives but
introduce resolution latency under mobility. ACP bridges agentic systems with legacy NF management interfaces across the full RAN--core--edge stack via a 
framework-agnostic REST interface. AP2 is confined to the business and service layers, supporting autonomous economic transactions through verifiable signed mandates in multi-operator ecosystems.
A cross-cutting constraint applies to all five protocols: JSON serialization introduces per-message payload overhead of 0.5--several~Kilobyte, making migration to binary encoding and persistent connection reuse a prerequisite for any
protocol targeting near-RT or RT 6G control tiers.

\subsection{Evaluation Methods}
Agentic AI systems introduce capabilities such as tool use, planning, multi-step reasoning, and autonomous decision-making, which extend beyond traditional language modeling benchmarks. Accordingly, evaluation must capture not only output correctness but also behavioral reliability, task efficacy, safety, and robustness in dynamic operational environments \cite{sec3.b37, sec3.b38}. We organize existing evaluation approaches into a compact taxonomy of metrics, while directing readers to prior works for detailed formulations due to space constraints.

\subsubsection{Task-Level Performance Metrics}

This category captures an agent’s ability to complete complex, multi-step workflows involving planning, tool invocation, and interaction with external systems. Representative metrics include task success rate, execution efficiency, error propagation, and constraint adherence. These measures are particularly relevant for telecom scenarios such as fault resolution, resource optimization, and closed-loop automation, with detailed treatments provided in \cite{sec3.b37}.

\subsubsection{Tool Interaction and Reliability Metrics}

Given the reliance of Agentic AI on external tools, APIs, and data sources, this category evaluates tool call correctness, argument validity, latency sensitivity, and recovery from failures. In network management contexts, such metrics reflect robustness in interfacing with telemetry systems, configuration platforms, and monitoring services. Comprehensive definitions and evaluation protocols are discussed in \cite{sec3.b38}.

\subsubsection{Efficiency and System-Level Metrics}

This category addresses the computational and operational overhead introduced by Agentic AI, including resource consumption, inference latency, and tool invocation cost. It further considers trade-offs between autonomy and system load, which are critical in telecom environments with strict real-time and energy constraints. Detailed analyses can be found in \cite{sec2.b14, sec3.b37, sec3.b38}.

\section{Agentic AI for 5G/6G Networks: A Survey}
\label{sec:agentic-for-ngn}

In this section, we review recent literature exploring LLM-based Agentic AI approaches for future networks, following the architectural framework established in Section~\ref{sec:ngn-tuto}. We organize prior work according to the three primary 6G layers illustrated in Fig.~\ref{fig:6g-layers}: (i) the \emph{control plane}, comprising the RAN, CN, TN, and the Edge/Cloud continuum; (ii) the \emph{management plane}, which focuses on frameworks enabling autonomous 6G operations such as IBN and ZSM; and (iii) the \emph{AI-native plane}, which encompasses the full lifecycle of agentic intelligence, including data operations, AI operations (training, inference, and orchestration), and AI explainability. Under this taxonomy, we analyze how Agentic AI optimizes 6G infrastructure resources, enables high-level closed-loop automation, and defines systematic methodologies for deploying intelligent and interpretable models across the network fabric. Collectively, these works illustrate a paradigm shift toward AI-native networking, where intelligence is no longer an overlay but a first-class architectural component of 6G systems. A structured summary of the surveyed contributions is provided in Table~\ref{tab:ai6g}.

\begin{table*}[t!]
\centering
\caption{Summary of LLM-based Agentic AI for 5G/6G Networks.}
\label{tab:ai6g}
\scriptsize
\renewcommand{\arraystretch}{1.1}
\setlength{\tabcolsep}{4pt}
\begin{tabular}{llp{2.2cm}p{5cm}p{3.0cm}}
\hline
\rowcolor{gray!15}
\textbf{Layer} & \textbf{Domain} & \textbf{Task} & \textbf{Focus} & \textbf{References} \\
\hline\hline

\multirow{17}{*}{\textbf{Control plane}}
& \multirow{6}{*}{RAN}
    & Resource Allocation
        & Agentic xApps/rApps for O-RAN control and RRM
        & \cite{elkael2025agentran,bao2025llm,salan2025rag,kamatani2025llm,feng2025towards} \\
    \cline{3-5}
    & & Security
        & Agentic and LLM-based threat detection and security compliance in O-RAN
        & \cite{chatzimiltis2025agentic,moore2025integrated,wen20246g} \\
    \cline{3-5}
    & & Configuration
        & Agentic observability platforms and conflict evaluation for xApps
        & \cite{sec6.b26,sharma2025towards,maxenti2025autoran} \\
    \cline{3-5}
    & & Intent-Based Control
        & LLM-as-operator and hierarchical RAN control
        & \cite{sec6.b27,elkael2025agentran,bao2025llm,salan2025rag} \\
    \cline{3-5}
    & & Simulation
        & Multi-agent LLM-driven RAN simulation and evaluation
        & \cite{rezazadeh2024genonet,hu2025reflection} \\
    \cline{3-5}
    & & Marketplace
        & Symbiotic/agentic marketplaces and safe coordination
        & \cite{chatzistefanidis2025symbiotic,zhang2025safe} \\
    \cline{2-5}

& \multirow{3}{*}{TN}
    & Configuration
        & LLM-driven SDN/SDM configuration automation
        & \cite{wang2024llm,neupane2025netprompt} \\
    \cline{3-5}
    & & Security
        & LLM-based intrusion and DDoS detection in SDN/SD-WAN
        & \cite{swileh2025proactive,lodh2025lightweight,zhang2023large,cao2019ai} \\
    \cline{3-5}
    & & Traffic Engineering
        & LLM-aware load balancing and graph-based dynamic networking
        & \cite{li2025load,sun2024large} \\
    \cline{2-5}

& \multirow{3}{*}{CN}
    & Intent Management
        & LLM-assisted intent extraction and semantic routing in 5G/6G
        & \cite{manias2024semantic,manias2024towards,rodriguez2024leveraging,kan2024mobile} \\
    \cline{3-5}
    & & Agentic Core Control
        & Mission-oriented and autonomous 6G core orchestration
        & \cite{tong2025core,li2025agentic,yu2025ai} \\
    \cline{3-5}
    & & LLM-Aware Slicing
        & Core/edge slicing tailored to LLM compute requirements
        & \cite{liu2024llm,maz2024llmxrl} \\
    \cline{2-5}

& \multirow{4}{*}{Edge/Cloud}
    & Service Placement
        & Agentic edge for AI services and semantic-aware offloading
        & \cite{tang2025end,feng2025towards,li2025netmcp,qian2024alibaba} \\
    \cline{3-5}
    & & Resource Allocation
        & Orchestration for AI-native and containerized applications
        & \cite{tang2025end,feng2025towards,kalafatidis2025llm,li2025netmcp} \\
    \cline{3-5}
    & & Security
        & Intrusion/threat detection for container environments
        & \cite{kalafatidis2025llm,rigaki2024hackphyr} \\
    \cline{3-5}
    & & Agent Economy
        & AI-native APIs enabling telco agent economies
        & \cite{barros2025ai,zhang2025safe} \\
\hline

\multirow{5}{*}{\textbf{Management plane}}
& \multirow{3}{*}{IBN}
    & Translation
        & Intent parsing and translation to network policies
        & \cite{mekrache2024intent,mekrache2024llm,mekrache2025oss,mekrache2025dmo,hossain2025netintent,angi2025llnet,lira2024large,ifland2024genet} \\
    \cline{3-5}
    & & Resolution
        & Intent conflict resolution and explainable justifications
        & \cite{salmi2025ai,ali2023leveraging} \\
    \cline{3-5}
    & & Assurance
        & End-to-end monitoring and SLA verification
        & \cite{tang2024large,mekrache2024combining,abbas2025ibn} \\
    \cline{2-5}

& \multirow{2}{*}{ZSM}
    & Service Automation
        & Zero-touch instantiation across TN/NTN segments
        & \cite{abbas2025ibn,mekrache2024combining} \\
    \cline{3-5}
    & & Closed-Loop Control
        & Multi-agent LLM architectures for self-organizing networks
        & \cite{qayyum2025llm,qu2025llm,gemayel2025network,shah2025tele} \\
\hline

\multirow{9}{*}{\textbf{AI-native plane}}
& \multirow{2}{*}{Data Ops.}
    & Data/Telemetry
        & NWDAF-style collection and log semanticization
        & \cite{quadrini2023data,rodriguez2024leveraging,kamatani2025llm,wu2024netllm,chochliouros2025developing} \\
    \cline{3-5}
    & & Datasets
        & Domain-specific QA and evaluation datasets for traffic data
        & \cite{ait20245g,wu2024netllm} \\
    \cline{2-5}

& \multirow{5}{*}{AI Ops.}
    & LLM Adaptation
        & Fine-tuning LLMs (Mobile-LLaMA) and context routing
        & \cite{ait20245g,kan2024mobile,wu2024netllm,li2025netmcp} \\
    \cline{3-5}
    & & Orchestration
        & Multi-agent design patterns for agentic RAN/Core/SON
        & \cite{mekrache2025oss,mekrache2025dmo,qayyum2025llm,qu2025llm,elkael2025agentran,li2025agentic,barros2025ai,maz2025eurocomply} \\
    \cline{3-5}
    & & Architecture
        & Blueprints for AI-native 6G and IBN-ZTSA integration
        & \cite{salmi2025ai,abbas2025ibn,feng2025towards,boutouchent20256g} \\
    \cline{3-5}
    & & Optimization
        & DRL-based task scheduling and inference optimization
        & \cite{mekrache2025drl,Fang2024} \\
    \cline{3-5}
    & & Evaluation
        & Benchmarking frameworks for LLMs and agents
        & \cite{maatouk2025teleqna,gupta2025mmtelco,colle2025telemath,ferrag2026alpha3,nikbakht2024tspec,wu2025tnautorca} \\
    \cline{2-5}

& \multirow{2}{*}{\parbox{1.8cm}{Explainability \\ \& Compliance}}
    & XAI
        & Causal inference and XAI for trustworthy control
        & \cite{sharma2025towards,mekrache2024combining,wen20246g,lodh2025lightweight} \\
    \cline{3-5}
    & & Compliance
        & Collaborative Agents for MLOps compliance with AI/Telecom regulation
        & \cite{maz2025eurocomply} \\
\hline

\end{tabular}
\end{table*}

\subsection{Control plane}

Agentic AI approaches are being extensively explored across all control plane components, spanning the RAN to the Edge/Cloud continuum. These works address critical challenges such as resource allocation, automated configuration, and security within each network domain. In the following, we review recent literature and developments specific to each of these areas.

\subsubsection{Radio Access Network}

In the RAN domain, AgentRAN introduces an AI-native, O-RAN-aligned agentic architecture in which LLM-powered agents interpret natural-language intents and orchestrate hierarchical control loops across rApps, xApps, and dApps for tasks such as scheduling and power control~\cite{elkael2025agentran}. This is complemented by ALLSTaR, which focuses on automated scheduler generation by synthesizing Medium Access Control (MAC) schedulers from intent-like specifications, and LLM-hRIC, which proposes a hierarchical RAN intelligent control framework~\cite{bao2025llm}. Furthermore, Giwa \emph{et al.} explore LLM-based assistants for intent-driven RAN management~\cite{sec6.b27}. Other proposals, such as RAG-empowered Radio Resource Management (RRM), LLM-xApp, and LLM-5GMAC, demonstrate how agentic reasoning improves spectral efficiency, observability, and troubleshooting in O-RAN deployments~\cite{salan2025rag,wu2025llm,kamatani2025llm}.

Beyond classical RRM, Agentic AI is increasingly explored for edge-native coordination and network assurance. Frameworks for Internet-of-Vehicles and semantic-aware 6G edge networks leverage agents to coordinate resource allocation across distributed nodes~\cite{feng2025towards}, while MAS handle traffic management and slice assurance in dense deployments~\cite{salama2025poster}. On the observability side, MX-AI introduces an agentic control platform for AI-RAN, and AutoRAN advocates zero-touch RAN operation through learning-driven automation~\cite{sec6.b26,maxenti2025autoran}. To address operational conflicts, recent work combines explainable machine learning and causal inference to diagnose and resolve conflicting control logic across xApps~\cite{sharma2025towards}.

Security and trustworthiness remain central challenges in agentic RAN design. Recent works include security-compliance agents, threat-mitigation frameworks such as MobiLLM, and secure slicing xApps for Open RAN~\cite{chatzimiltis2025agentic,moore2025integrated,wen20246g}. To ensure safe agent interactions in open marketplaces, Zhang \emph{et al.} propose frameworks for secure multi-agent coordination~\cite{zhang2025safe}. Additionally, reflection-driven self-optimization and symbiotic agent models are proposed to enhance trustworthiness in AGI-driven RANs~\cite{hu2025reflection,chatzistefanidis2025symbiotic}. Finally, simulation platforms such as GenoNet enable multi-agent RAN experimentation using ns-3, while Tele-LLM-Hub and NetMCP provide operator assistants and network-aware context protocols for RAN controllers~\cite{rezazadeh2024genonet,shah2025tele,li2025netmcp}.

\subsubsection{Transport Network}

In the TN, agentic controllers are primarily deployed for configuration automation, security enforcement, and traffic engineering. LLM-enabled full-stack configuration frameworks for Space Division Multiplexing (SDM) networks and LLM-driven SDN controllers for optical transport translate high-level intents into device-level configurations, significantly simplifying the management of complex IP and optical infrastructures~\cite{wang2024llm}. NetPrompt extends this paradigm by synthesizing SDN policies from operator directives, while graph-based LLM frameworks abstract complex path computation and traffic engineering decisions~\cite{neupane2025netprompt,sun2024large}.

Security in TNs is addressed through LLM-based intrusion and anomaly detection in SDN and Software-Defined Wide Area Network (SD-WAN) environments. This includes lightweight fine-tuning for explainable intrusion detection, zero-training proactive Distributed Denial of Service (DDoS) mitigation using port-level monitoring, and malicious packet recognition~\cite{lodh2025lightweight,swileh2025proactive}. Earlier agent-based models laid the foundation for network service prediction and resource scheduling, while contemporary LLM-driven approaches focus on intelligent SD-WAN maintenance, including alarm correlation and root-cause analysis~\cite{cao2019ai,zhang2023large}. Finally, agentic load-balancing mechanisms leverage LLM-based reasoning to optimize flow distribution across dynamic transport topologies~\cite{li2025load}.

\subsubsection{Core Network}

In the CN, semantic and agentic control frameworks are being developed to realize intent-based and mission-oriented operations. Semantic routing mechanisms leverage LLMs to map natural-language intents into declarative policies that steer traffic through appropriate NF chains~\cite{manias2024semantic}. This direction is further supported by research on LLM-based intent extraction and trace analytics for proactive core monitoring~\cite{manias2024towards,rodriguez2024leveraging}. Mobile-LLaMA extends this capability through instruction fine-tuning of open-source models, enabling precise analysis of GTP protocols, control-plane messaging, and KPIs for rapid troubleshooting~\cite{kan2024mobile}.

Agentic AI has also been proposed to redefine core orchestration. Frameworks such as A-Core and Agentic-AI Core introduce mission-oriented agentic controllers that coordinate core functions, while autonomous cognitive architectures rely on collaborative agents to manage mobility, session continuity, and policy enforcement~\cite{tong2025core,li2025agentic,yu2025ai}. Moreover, LLM-Slice introduces dedicated network slicing for LLM traffic by jointly considering compute and network requirements, directly coupling agentic application needs with core resource provisioning~\cite{liu2024llm,maz2024llmxrl}.

\subsubsection{Edge/Cloud}
At the Edge/Cloud layer, Agentic AI primarily targets AI-service hosting, resource management, and cloud-native security. End-to-end edge AI provisioning frameworks orchestrate AI workloads across RAN and edge segments in 6G O-RAN environments, while semantic-aware edge networks dynamically determine optimal model placement and task offloading under latency and energy constraints~\cite{tang2025end,feng2025towards}. NetMCP facilitates this integration by introducing a network-aware MCP that enables LLMs to discover and interact with network APIs for fine-grained resource control~\cite{li2025netmcp}. Large-scale deployments, such as Alibaba HPN, further demonstrate data-center networks optimized for LLM training and inference traffic patterns~\cite{qian2024alibaba}.

Security in distributed Edge/Cloud environments is increasingly handled by specialized agents. Recent works propose LLM-enhanced intrusion detection for containerized applications using multi-tier IDS architectures for SDN and Kubernetes~\cite{kalafatidis2025llm}. Similarly, Hackphyr explores locally fine-tuned LLM agents for secure network operations in edge and on-premise settings~\cite{rigaki2024hackphyr}. Finally, the emergence of a ``telco agent economy'' is supported by AI-native network APIs that enable external agents to safely invoke, compose, and monetize network services over programmable Edge/Cloud infrastructures~\cite{barros2025ai,zhang2025safe}.

\noindent\textbf{Synthesis and Takeaway Lessons.}
The surveyed literature collectively evidences a transition toward intent-driven, agentic control across RAN, transport, core, and edge domains, yet converges on a consistent set of unresolved system-level constraints. The main conclusions are as follows: (i) there is a fundamental latency–accuracy trade-off, as LLM-based reasoning cannot satisfy sub-10 ms control-loop requirements, which necessitates hierarchical decomposition across non-RT and near-RT planes with model compression or speculative execution, albeit introducing additional and insufficiently quantified E2/A1/O1 signaling overhead; (ii) intent-to-policy translation is fragile under topology changes and KPI variability, requiring continuous re-grounding mechanisms closely integrated with real-time observability; (iii) embedding agents within the managed infrastructure creates circular resource dependencies that complicate system stability and resource allocation.

\subsection{Management plane}

The management plane encompasses Agentic AI frameworks that operate above the raw infrastructure, with a primary focus on IBN and ZSM. In this layer, LLMs and MAS are leveraged to close the loop between high-level operator goals and low-level network configurations, as well as to automate service lifecycle operations across heterogeneous domains.

\subsubsection{Intent-Based Networking}
A primary group of works focuses on the \emph{intent translation problem}, which consists of mapping natural-language goals into machine-executable policies. Mekrache \emph{et al.} propose several LLM-centric IBN pipelines in which intents are parsed, semantically structured, and mapped to network actions, with a particular emphasis on next-generation softwarized networks, Operations Support Systems (OSS), and Business Support Systems (BSS) integration~\cite{mekrache2024intent,mekrache2024llm,mekrache2025oss,mekrache2025dmo}. Complementarily, NetIntent introduces an end-to-end SDN framework and the IBNBench dataset for benchmarking LLMs on intent translation and flow-conflict detection while coordinating agents across ONOS controllers~\cite{hossain2025netintent}. At the device level, LLNeT employs a Small Language Model (SLM) to directly instruct softwarized network elements, whereas Genet provides multimodal interfaces that translate high-level descriptions into low-level configurations, reducing operational errors~\cite{angi2025llnet,ifland2024genet,lira2024large}.

Beyond translation, recent literature addresses \emph{intent conflict resolution}. AI-native O-RAN architectures for 6G incorporate intent-aware control mechanisms to reconcile competing objectives, such as throughput versus energy efficiency~\cite{salmi2025ai}. Similarly, LLM-assisted Deep Reinforcement Learning (DRL) strategies have been proposed for anti-jamming, where agents iteratively refine policies under adversarial conditions~\cite{ali2023leveraging}.

A third stream of work focuses on \emph{intent assurance and monitoring}. LLM-assisted health management architectures perform semantic analysis of telemetry to verify whether deployed configurations satisfy the original intents~\cite{tang2024large}. Mekrache \emph{et al.} further combine XAI and LLMs to provide human-understandable justifications for automated decisions across the IBN pipeline, thereby increasing trust in closed-loop adaptations~\cite{mekrache2024combining}. This vision is extended by IBN-ZTSA frameworks, which integrate intent-based logic with zero-touch automation for continuous SLA verification in hybrid Terrestrial and Non-Terrestrial Networks (TN/NTN)~\cite{abbas2025ibn}.

\subsubsection{Zero-touch network and Service Management}

In the ZSM domain, Agentic AI is utilized to realize self-organizing, self-healing, and self-optimizing networks across the RAN, CN, TN, and Edge/Cloud. LaMA-SON introduces an LLM-driven multi-agent architecture for intelligent Self-Organizing Network (SON) management, in which specialized agents handle traffic management, QoS optimization, and security threat detection using role-specific reasoning~\cite{qayyum2025llm}. This multi-agent paradigm is further extended by Qu \emph{et al.}, who propose a dual-loop edge--terminal collaboration framework for 6G that coordinates agents at both the network edge and user terminals to jointly optimize radio and computing resources~\cite{qu2025llm}.

At the orchestration plane, LLM-based MAS automates the lifecycle of Virtualized Network Functions (VNFs) by mapping service descriptions into NF graphs and triggering scaling operations when needed~\cite{gemayel2025network}. Tele-LLM-Hub generalizes this concept by providing a context-aware platform where agents collaborate via the TeleMCP protocol to support fault management and performance analysis across domains~\cite{shah2025tele}.

Finally, zero-touch service automation is explicitly addressed in the IBN-ZTSA framework, which couples intent-based pipelines with ZSM to automate service instantiation and scaling across TN/NTN segments~\cite{abbas2025ibn}. Complementing this, Mekrache \emph{et al.} demonstrate that combining XAI and LLMs embeds trust and interpretability into the ZSM loop, enabling autonomous yet auditable scaling and anomaly resolution~\cite{mekrache2024combining}.

\begin{table*}[t]
\centering
\caption{Comparison of Agentic AI Approaches in the Management plane.}
\label{tab:agentic_management_telecom}
\scriptsize
\renewcommand{\arraystretch}{1.0}
\setlength{\tabcolsep}{3pt}
\begin{tabular}{p{0.7cm}p{2.4cm}p{2.5cm}p{2.5cm}p{1.8cm}p{1.8cm}p{2.5cm}}
\hline
\rowcolor{gray!15}
\textbf{Work} & \textbf{Agentic Arch.} & \textbf{LLM} & \textbf{Memory} & \textbf{Tools} & \textbf{Domain} & \textbf{KPI} \\
\hline\hline

\cite{mekrache2024intent}
& Single LLM (ReAct)
& Ollama (Mistral:7B, Llama:13B)
& Ext. KB + Prompt Eng.
& OSS API
& Cloud-Edge/RAN
& Avg. score, Execution time \\
\hline

\rowcolor{gray!15}
\cite{mekrache2025oss}
& MAS (Hierarchical)
& Finetuned LLM + GPT-4
& Ext. KB + Prompt Eng.
& OSS API
& Cloud-Edge/RAN/Core
& BERT, Cosine, Exact Match \\
\hline

\cite{hossain2025netintent}
& MAS-Hybrid (LLM/Non-LLM)
& Ollama (Qwen, Phi, Llama, etc.)
& Prompt Eng. (short-term)
& SDN (ONOS)
& SDN
& Accuracy, Execution time \\
\hline

\rowcolor{gray!15}
\cite{angi2025llnet}
& Single LLM (ReAct)
& Gemini-1-Pro, Phi3-mini, Llama3:8B
& Prompt Eng. (short-term)
& SDN (Ryu)
& SDN
& Latency, Accuracy, Energy, Token, Cost \\
\hline

\cite{ifland2024genet}
& Single Multimodal LLM
& GPT-4-Vision
& Prompt Eng. (short-term)
& Network Topology
& Net. Configuration
& Human/LLM Correlation \\
\hline

\rowcolor{gray!15}
\cite{lira2024large}
& Single LLM (ReAct)
& Ollama (zephyr:7b)
& Prompt Eng. (short-term)
& Network Topology
& Net. Configuration
& Accuracy, Execution time \\
\hline

\cite{salmi2025ai}
& MAS (Hierarchical)
& Not Mentioned
& Ext. Memory + Prompt Eng.
& O-RAN xApps/rApps
& RAN
& No Evaluation \\
\hline

\rowcolor{gray!15}
\cite{ali2023leveraging}
& MAS-Hybrid (DRL-LLM)
& Ollama (Falcon)
& Prompt Eng. (short-term)
& RL environment
& RAN
& Policy reward signal \\
\hline

\cite{maz2025eurocomply}
& MAS (Hierarchical)
& Ollama (Mistral, Qwen, deepseek-r1, etc.)
& Agentic RAG (long+short)
& Regulation API, Web Search
& Core/RAN/Edge/ Compliance
& Resp. time, Score, LLM-as-Judge \\
\hline

\rowcolor{gray!15}
\cite{qayyum2025llm}
& MAS (Custom)
& Ollama (Llama, Qwen, etc.)
& Cloud memory (centralized)
& Network management
& Self-Org. Networks
& Accuracy, Score rating \\
\hline

\cite{qu2025llm}
& MAS (Distributed)
& MiniCPM-V2.6
& Ext. Memory + Prompt Eng.
& Map, Meteorology APIs
& Urban Emergency
& Success rate, Avg. delay \\
\hline

\rowcolor{gray!15}
\cite{gemayel2025network}
& MAS (CoT)
& GPT-2, RoBERTa, etc.
& RAG
& rApp integration
& RAN
& Accuracy \\
\hline

\cite{ameur2026agentic}
& MAS (Hierarchical)
& Ollama (gpt-oss, gemma3, deepseek-r1, etc.)
& Agentic RAG + MCP
& MCP (KB + NWDAF)
& Core (NWDAF)
& Task/Tool accuracy, LLM-as-Judge \\
\hline

\end{tabular}
\end{table*}

\subsubsection*{\textbf{Agentic Capabilities for the Management plane}}

To ensure both depth and clarity, we deliberately chose to expand our focus to the management layer, which represents the most immediate and impactful domain for integrating LLM-based agentic frameworks. This layer inherently aligns with the strengths of Agentic AI, including decision-making, orchestration, policy adaptation, and closed-loop automation across complex and dynamic network environments. Moreover, this focus is strongly motivated by the fact that contemporary telecom standardization efforts, such as those led by the ETSI within the ETSI GR ENI framework, explicitly emphasize intelligence and automation at the management and orchestration layers. These initiatives highlight the centrality of this layer in enabling cognitive network operations and validating AI-driven control loops.

Table~\ref{tab:agentic_management_telecom} synthesizes representative LLM-based agentic frameworks for NGN management along the axes of architectural paradigm, model choice, tool ecosystem, memory design, and reported KPIs. The table reveals a clear progression from monolithic LLM agents toward distributed, hierarchical, and hybrid MAS, driven by the scalability, robustness, and control-loop requirements of telecom deployments.
Early works such as \cite{mekrache2024intent} and \cite{lira2024large} adopt a single-agent ReAct paradigm in which reasoning and acting are co-located within one LLM instance, typically a GPT-4-class model selected for its zero-shot reasoning and tool-use capabilities. Their evaluation centers on semantic accuracy, intent translation fidelity, and execution latency. While these models deliver strong reasoning precision, they are constrained by bounded context windows, the absence of persistent memory, and prompt sensitivity, which limit reproducibility and robustness in multi-domain scenarios.
Hierarchical MAS frameworks, including \cite{mekrache2025oss}, \cite{maz2025eurocomply}, and \cite{ameur2026agentic}, mark a substantive architectural shift. They deploy specialized agents backed by heterogeneous LLMs, combining fine-tuned domain-specific models with high-capacity proprietary ones to match reasoning depth and latency to task requirements. Evaluation moves beyond isolated accuracy toward system-level indicators such as SLA compliance, orchestration latency, and cross-domain coordination efficiency. Agentic RAG and hybrid memory mechanisms strengthen contextual grounding across iterative control loops, and the emerging LLM-as-a-judge paradigm enables meta-evaluation that partially compensates for the lack of standardized telecom benchmarks.

Hybrid architectures, exemplified by \cite{hossain2025netintent} and \cite{ali2023leveraging}, separate high-level reasoning from low-level control execution. LLMs handle intent interpretation, policy abstraction, and decision guidance, while deterministic controllers such as SDN frameworks or DRL agents perform real-time optimization. This decoupling improves convergence time, policy optimality, and stability in dynamic RAN environments. The LLM–DRL coupling is particularly illustrative: the LLM constrains the policy search space while DRL provides adaptability and performance guarantees under stochastic conditions \cite{maz2024llmxrl}, addressing the lack of formal convergence properties in purely generative approaches.

Two transversal dimensions further differentiate these systems. First, \textit{tool ecosystems} have evolved from narrow OSS/BSS and SDN bindings toward broader integration with O-RAN rApps/xApps, NWDAF analytics, and external data sources, with standardized protocols such as MCP \cite{ameur2026agentic} enabling structured and interoperable tool invocation. This repositions the LLM as an orchestrator of tool-augmented cognition rather than a standalone decision engine, at the cost of new overheads in tool invocation latency and interoperability. Second, \textit{memory design} has progressed from stateless prompt-based context toward multi-layered architectures combining short-term conversational state with long-term knowledge persistence via agentic RAG or centralized repositories. These designs improve long-horizon task performance but introduce consistency, synchronization, and retrieval-efficiency challenges, particularly in distributed deployments such as \cite{qu2025llm}.


\noindent\textbf{Synthesis and Takeaway Lessons.}
At the management plane, Agentic AI strengthens the coupling between high-level intents and automated execution across IBN and ZSM paradigms, but several limitations persist. The main conclusions are as follows: (i) intent translation fidelity degrades under ambiguity, multi-domain interactions, and conflicting policies, due to context limitations and the absence of consistency-preserving mechanisms; (ii) hierarchical multi-agent systems improve SLA alignment and cross-domain coordination, but introduce orchestration and inter-agent communication overhead that is rarely benchmarked against ZSM latency constraints; (iii) hybrid LLM–DRL approaches accelerate convergence, yet lack formal stability guarantees under stochastic dynamics; (iv) tool ecosystems remain heterogeneous across OSS/BSS, SDN, O-RAN, and NWDAF, with invocation latency and interoperability costs largely unreported, while memory architectures trade temporal coherence against synchronization and retrieval overhead; (v) the tight integration of semantic intent handling with autonomous orchestration creates compounded failure modes, where misinterpretations propagate across control loops in the absence of explicit validation checkpoints.

\subsection{AI-native plane}

The AI-native plane encompasses the full lifecycle of agentic intelligence within the network, spanning data collection and semanticization (Data Operations), model adaptation and multi-agent orchestration (AI Operations), and mechanisms for ensuring trustworthy, interpretable deployments (Explainability). This holistic perspective positions Agentic AI as a native architectural component of the 6G ecosystem rather than an external add-on.

\subsubsection{Data Operations}

Several works focus on the data plane that feeds Agentic AI. Quadrini \emph{et al.} demonstrate how the 3GPP NWDAF can be used to collect real 5G core traffic for analytics and downstream AI tasks, providing a practical blueprint for data acquisition in operational networks~\cite{quadrini2023data}. Building on this, Rodriguez-Navas \emph{et al.} propose LLM-assisted trace analytics for next-generation mobile cores, in which traces are ingested, semantically transformed, and queried via LLMs to support troubleshooting and performance analysis~\cite{rodriguez2024leveraging}.

At the RAN side, Kamatani \emph{et al.} leverage LLMs to analyze MAC-layer logs in O-RAN split 7.2, semantically summarizing log sequences and identifying performance bottlenecks that are difficult to capture with rule-based tools~\cite{kamatani2025llm}. NetLLM generalizes this idea through a multimodal encoder and networking heads that transform heterogeneous inputs such as time series, logs, and topologies, into unified token-like embeddings, enabling a single model to handle diverse networking tasks~\cite{wu2024netllm}. Chochliouros \emph{et al.} further emphasize the importance of telecom-specific dataset design and data quality for LLM training in AI-native networks~\cite{chochliouros2025developing}.

On the evaluation side, TrafficNetQA introduces question-answering datasets tailored to traffic network files, enabling systematic assessment of an LLM's ability to interpret spatial-temporal traffic data~\cite{kwon2025trafficnetqa}. Similarly, Abderrahmane \emph{et al.} construct 5G-focused datasets for tasks such as KPI forecasting and anomaly detection, bridging the gap between generic benchmarks and telecom-specific distributions~\cite{ait20245g}.

\subsubsection{AI Operations}
AI Operations (AIOps) focus on how models and agents are adapted, orchestrated, and optimized within 6G systems. Mobile-LLaMA applies instruction fine-tuning to open-source LLMs so they can interpret 5G logs and KPIs, demonstrating that domain adaptation significantly improves alarm classification and root-cause analysis~\cite{kan2024mobile}. NetLLM proposes a Low-Rank Networking Adaptation (DD-LRNA) scheme that enables a base LLM to solve a wide range of networking tasks with modest overhead~\cite{wu2024netllm}. NetMCP further extends this capability by defining a network-aware MCP and the SONAR routing algorithm, which jointly consider semantic similarity and real-time QoS when invoking external tools~\cite{li2025netmcp}.

Multi-agent orchestration is studied at both the management plane and within network architectures. Qayyum \emph{et al.} design LLM-driven multi-agent SON frameworks where specialized agents handle traffic engineering and fault management~\cite{qayyum2025llm}. Qu \emph{et al.} propose a dual-loop edge--terminal system in which terminal-side and edge-side agents collaborate to optimize radio and compute resources~\cite{qu2025llm}. Architecturally, AgentRAN and Agentic-AI Core embed hierarchies of agents directly into RAN and core designs, enabling mission-oriented control loops~\cite{elkael2025agentran,li2025agentic}. Barros introduces AI-native network APIs for a ``telco agent economy,'' defining how external agents can safely monetize network capabilities~\cite{barros2025ai}.

Several works advocate fully AI-native 6G architectures. AI-native O-RAN designs assume AI components as first-class citizens in both control and data planes~\cite{salmi2025ai}. IBN-ZTSA extends this philosophy to end-to-end service automation across TN/NTN domains~\cite{abbas2025ibn}. Feng \emph{et al.} propose 6G native-AI edge networks where semantic intelligence is tightly coupled with task-oriented communication~\cite{feng2025towards}, while Boutouchent \emph{et al.} outline high-level principles for placing AI functions across the RAN--core--cloud continuum~\cite{boutouchent20256g}.

Runtime optimization is also a key research direction. Mekrache \emph{et al.} introduce DRL-based schedulers for LLM workloads that dynamically allocate GPU and network resources~\cite{mekrache2025drl}, while Fang \emph{et al.} study DRL-controlled inference pipelines that decide between edge and cloud execution~\cite{Fang2024}. In parallel, a growing ecosystem of benchmarks (summarized in Table \ref{tab:telecom_llm_benchmarks}), TeleQnA, TeleTables, TelAgentBench, MMTelCo, TeleMath, TSpec-LLM, $\alpha$3-Bench, and TeleYAML, enables systematic evaluation of telecom-focused LLMs and agents~\cite{maatouk2025teleqna,gupta2025mmtelco,colle2025telemath,ferrag2026alpha3,nikbakht2024tspec,wu2025tnautorca, gsmabenchmarks}.

\begin{table*}[t]
\centering
\caption{Summary of Telecom-Oriented Agentic AI Benchmarks.}
\label{tab:telecom_llm_benchmarks}
\small
\renewcommand{\arraystretch}{1.0}
\setlength{\tabcolsep}{4pt}
\rowcolors{2}{gray!15}{white}
\begin{tabular}{p{2.2cm} p{3.0cm} p{3.0cm} p{5.5cm}}
\hline
\rowcolor{gray!15}
\textbf{Benchmark} & \textbf{Core Task} & \textbf{Metric} & \textbf{Benchmark Focus} \\
\hline\hline

TeleQnA       & Telecom Knowledge QA      & Exact-Match Accuracy         & Multiple-choice questions covering telecom standards, terminology, and domain knowledge. \\ \hline
TeleTables    & Table Interpretation      & Exact-Match / MCQ Accuracy   & Evaluation of LLM ability to understand and reason over telecom standard tables. \\ \hline
TelAgentBench & Agentic Telecom Eval.     & Multi-capability Metrics     & Benchmark for agentic capabilities (reasoning, planning, tool use, RAG, instruction following) in telecom tasks. \\ \hline
MM-Telco      & Multimodal Telecom Tasks  & Mixed (MCQ, retrieval)       & Suite of multimodal benchmarks involving text/image tasks for telecom use cases. \\ \hline
TeleMath      & Mathematical Reasoning    & Exact-Match                  & Telecom-specific math problem solving and numerical reasoning. \\ \hline
TSpec-LLM     & Standards Comprehension   & Exact-Match / Retrieval      & Comprehensive dataset for LLM interpretation of 3GPP technical specifications. \\ \hline
TeleYAML      & Intent-to-Config. Gen.    & Graded Score                 & Generation of structured telecom config from natural language intent (GSMA Open-Telco task). \\ \hline
$\alpha$3-Bench & Conversational UAV Autonomy & Composite Score           & Benchmarking safe, network-aware, and resource-efficient LLM agents in 6G-enabled aerial systems. \\
\hline

\end{tabular}
\end{table*}

\subsubsection{Explainability \& Compliance} Explainability is essential to ensure that Agentic AI decisions remain transparent, auditable, and trustworthy. In the RAN, Sharma \emph{et al.} combine explainable ML and causal inference to diagnose and mitigate xApp conflicts~\cite{sharma2025towards}. 6G-xSec introduces an explainable edge security framework for Open RAN that clarifies why specific flows are flagged as malicious~\cite{wen20246g}. At the management level, Mekrache \emph{et al.} advocate the joint use of XAI and LLMs to support human-in-the-loop validation within ZSM~\cite{mekrache2024combining}. In transport security, Lodh \emph{et al.} show that feature-attribution explanations can help operators calibrate trust in LLM-based SDN intrusion detection systems~\cite{lodh2025lightweight}. Beyond explainability, regulatory compliance is an emerging concern: EUROCOMPLY proposes an agentic MLOps framework that automates compliance verification against AI and telecom regulations, embedding policy-aware agents directly into the model lifecycle~\cite{maz2025eurocomply}.

\noindent\textbf{Synthesis and Takeaway Lessons.}
At the AI-native plane, the viability of agentic 6G systems is constrained by tightly coupled challenges across data pipelines, model orchestration, and runtime deployment. The main conclusions are as follows: (i) telecom-specific DataOps pipelines emphasize semanticization and quality control, but the lack of standardized interfaces between NWDAF outputs and LLM-ready representations leads to ad hoc preprocessing, limiting reproducibility and benchmark portability; (ii) in AI Operations, multi-agent decomposition, model adaptation, and DRL-based scheduling enable functional scalability, yet inter-agent communication overhead, consensus latency, and tool-routing delays remain unquantified relative to ZSM and SON timing constraints, while inference optimization is not aligned with network control objectives, leaving end-to-end latency budgets unmet; (iii) AI-native architectures distributing inference across the RAN–core–cloud continuum introduce significant context migration and session state transfer costs that scale with model size and interaction history; (iv) explainability is treated as a post-hoc process and does not provide corrective feedback into agent decision loops, limiting its utility for real-time control; (v) despite emerging telecom-specific benchmarks, evaluation remains fragmented and lacks standardization for consistent performance assessment.

\section{Standardization and Projects}
\label{sec:standards-projects}

This section reviews key standardization activities and collaborative research projects that explore the integration of LLM-based Agentic AI into NGN architectures.

\subsection{Standardization Efforts}

Standardization plays a foundational role in shaping the evolution of the telecommunications industry by translating emerging research concepts into interoperable, scalable, and deployable technologies. By defining common architectures, interfaces, and operational principles, standards enable multi-vendor adoption of AI-driven automation with predictable behavior and trust. As the industry moves toward 6G, Standards Development Organizations (SDOs) are formalizing LLM-based agents as native architectural components of network control, management, and orchestration, as summarized in Table~\ref{tab:agentic-standards}.

\begin{table*}[t]
\centering
\caption{Key Standardization and Industry Efforts Relevant to LLM-Based Agentic AI in 6G Networks.}
\label{tab:agentic-standards}
\scriptsize
\rowcolors{2}{gray!10}{white}
\renewcommand{\arraystretch}{1.0}
\setlength{\tabcolsep}{4pt}
\begin{tabular}{p{1cm}p{3.0cm}p{3.8cm}p{3.8cm}p{3.3cm}}
\hline
\rowcolor{gray!25}
\textbf{Org.} &
\textbf{Scope and Key Docs} &
\textbf{Agentic AI Technical Focus} &
\textbf{6G Integration Point} &
\textbf{Open Technical Gap} \\
\hline\hline

3GPP &
6G service/architecture; TR~22.870, SA2 AI-native studies &
LLM agents as native 6G NFs; intent-aware SBA interfaces; multi-agent service coordination across network domains. &
6G Core SBA (N-series interfaces); NRF-based NF registration extended to agent identities. &
LLM agent identity and OAuth2 token binding for ephemeral agent instances not yet specified. \\ \hline

ETSI &
ZSM ISG (GS ZSM 020/022); ENI ISG (GR~051, GR~055, GR~056, GS~059, GR~060--062); NFV ISG &
Closed-loop intent realization via ZSM management domains; ENI GR~056 multi-agent coordination for core networks; GS~059 agent interface and protocol specification (telecom-native agentic protocol); GR~062 security aspects; NFV MANO intent interfaces. &
E2E management fabric; cross-domain intent interfaces; MANO orchestration workflows; GS~059 agent interaction primitives for 5G/6G core. &
No binding latency requirements for closed-loop intent realization; GS~059 lacks wire-level serialization format, leaving 3GPP N-series and TM Forum API interoperability implementation-dependent. \\ \hline

TM Forum &
Autonomous Networks program; TMF Open APIs (TMF620/641/688); ODA &
Agent-driven OSS/BSS workflows; LLM intent decomposition over Open APIs; model-agnostic A2A protocol for inter-agent coordination; ODA AI/data capabilities. &
OSS/BSS northbound; TMF Open API tool invocation via MCP-compatible function-call schemas. &
A2A protocol lacks binding serialization format; wire-level interoperability with 3GPP/ETSI intent interfaces underspecified. \\ \hline

ITU-T &
IMT-2030 (Focus Group AI-native Networks, SG13); X.700-series alignment &
AI agents as perception-reasoning-decision-coordination functional entities; LLMs for intent reasoning over X.700-series management interfaces; agent lifecycle management. &
AI-native management plane; supervisory agent trust frameworks with configurable autonomy thresholds. &
Agent lifecycle procedures for stateful LLM sessions not yet harmonized with 3GPP NF lifecycle management. \\ \hline

O-RAN Alliance &
NG-RG 6G studies; O-RAN WG2 (non-RT RIC/A1); WG3 (near-RT RIC/E2) &
LLM agents as rApps (non-RT RIC, A1 intent policies) and xApps (near-RT RIC, E2 control, 10--100\,ms loop); intent mediation, policy planning, adaptive RAN control. &
Non-RT RIC (rApp, A1/O1 interfaces); near-RT RIC (xApp, E2 interface); open fronthaul for edge inference. &
No certification path for distilled LLM xApps at near-RT RIC; latency-accuracy trade-off at E2 boundary unresolved. \\ \hline

IETF &
IETF 124 Proceedings and  draft\_agentic\_usecases
 &
Protocol-level agent capability discovery and context sharing; YANG model extensions for agent-readable network state; delegated OAuth~2.0 credentials for agent authorization. &
Northbound RESTCONF/NETCONF interfaces; multi-operator agent identity federation. &
Delegated credential binding for ephemeral LLM agents across multi-operator domains lacks standardized solution. \\ \hline

GSMA &
Open-Telco LLM Benchmarks; TelecomGPT; Foundry pilots &
Quantitative benchmarking of LLM/agent capabilities on 3GPP spec comprehension, KPI-driven configuration, fault diagnosis; end-to-end agent workflow evaluation covering latency, energy, and API overhead. &
Operator validation layer bridging research benchmarks and 3GPP/ETSI deployment requirements. &
Benchmark KPI sets incompatible with 3GPP/ETSI metrics; unified cross-SDO evaluation framework absent. \\
\hline

\end{tabular}
\end{table*}

At the service and core architecture layer, 3GPP TR~22.870 envisions AI agents as first-class entities within the 6G SBA, interpreting user intents and coordinating services across domains through intent-aware NF interfaces~\cite{sec6.b1,sec6.b2}. The International Telecommunication Union Telecommunication Standardization Sector (ITU-T) Focus Group on AI-native Networks (SG13) complements this by specifying perception–reasoning–decision–coordination roles for agents and aligning LLM-driven intent reasoning with X.700-series management interfaces, while mandating human-in-the-loop validation above configurable autonomy thresholds~\cite{sec6.b31,sec6.b33}. 
Management and orchestration frameworks are addressed by ETSI, whose ZSM ISG defines closed-loop, cross-domain intent interfaces~\cite{sec6.b4,sec6.b5} and whose ENI ISG has produced the most mature telecom-native agentic protocol stack to date, spanning multi-agent coordination (GR~056), the agent interface specification (GS~059), and security aspects of AI agent-based cores (GR~062), with complementary integration into NFV-MANO workflows~\cite{sec6.b6,sec6.b7}.

TM Forum operationalizes these principles in OSS/BSS through its five-level Autonomous Networks model, in which agents invoke TMF620/641/688 Open APIs as tools via MCP-compatible function-call schemas and coordinate through a model-agnostic A2A protocol~\cite{sec6.b14,sec6.b16,sec6.b18}. At the RAN, the O-RAN Alliance maps LLM agents onto its disaggregated RIC architecture, deploying them as rApps consuming A1 policy intents at the non-RT tier and, where latency permits, as distilled xApps acting on E2 indications within the 10--100\,ms control loop~\cite{sec6.b23,sec6.b24,sec6.b25,sec6.b26}. Internet Engineering Task Force (IETF) provides the underlying protocol substrate through use-case-driven requirements for agent capability discovery and context sharing, alongside YANG/RESTCONF extensions and OAuth~2.0-based delegated credentials for agent authorization~\cite{ietf_agentic_ai_blog_2024,ietf_aiproto_usecases_2024}. Finally, Global System for Mobile communications (GSMA) anchors these efforts empirically: its Open-Telco LLM Benchmarks quantify the performance gap between general-purpose and telecom fine-tuned models (e.g., TelecomGPT) on 3GPP comprehension and KPI-driven configuration tasks, and extend evaluation to end-to-end agent workflows covering latency, energy, and API overhead~\cite{sec6.b35,sec6.b36,sec6.b38}.

Across all SDOs, three open gaps recur and are detailed per-organization in Table~\ref{tab:agentic-standards}: (i) identity and credential binding for ephemeral LLM agents, unresolved in both 3GPP NRF/OAuth2 procedures and IETF delegated-credential schemes; (ii) wire-level interoperability between ETSI GS~059, TM Forum A2A, and 3GPP N-series interfaces, which currently lack a shared serialization format; and (iii) the latency–accuracy trade-off for agent placement, most acute at the O-RAN near-RT RIC boundary and compounded by the absence of unified cross-SDO benchmarking KPIs.

\subsection{6G Project Initiatives}

Beyond formal standardization, large-scale 6G research initiatives validate AI-native architectural concepts through experimentation and prototyping, with LLM-based Agentic AI emerging as a recurring design primitive for translating service objectives into coordinated cross-domain actions. Table~\ref{tab:agentic-6g-projects} summarizes the representative initiatives discussed below, detailing their agentic architectures, integration points, and current limitations. Within the European SNS-JU programme, four complementary projects explore distinct layers of the agentic stack. SUNRISE-6G deploys LLM agents as intent mediators that translate natural-language service requests into platform-specific orchestration primitives over TMF-aligned northbound APIs, establishing Agentic AI as a federation layer across heterogeneous 6G testbeds~\cite{sec6.b39}.

6G-INTENSE advances this toward full Intent-Based Networking (IBN) lifecycle management, orchestrating a multi-agent pipeline that decomposes intents into domain sub-intents, verifies feasibility against real-time network state, enforces policies, and closes the assurance loop through KPI monitoring aligned with ETSI ZSM and TM Forum TMF921 interfaces~\cite{sec6.b40}. 6G-DALI extends agentic control into the data and model-lifecycle plane, using LLM agents to orchestrate DataOps and MLOps workflows, including dataset generation, preprocessing, training scheduling, and reproducibility, across NWDAF-compatible nodes~\cite{sec6.b41}. FLECON-6G complements these with a Network Digital Twin (NDT) substrate over which agents query twin state to compute cross-domain control updates, with explainability enforced through causal justification logs traceable to originating intents~\cite{sec6.b42}. Beyond Europe, the U.S. DoD-sponsored OPEN6G initiative targets the RAN tier through its AgentRAN hierarchical multi-agent system, deploying LLM rApps that consume A1 intent policies at the non-RT RIC and lightweight inference xApps that act on E2 indications within the 10--100\,ms control loop, with digital-twin validation gating live RAN commitment~\cite{sec6.b44}.

Collectively, these projects converge on three common limitations detailed per-project in Table~\ref{tab:agentic-6g-projects}: the absence of formal schema alignment and convergence guarantees for intent grounding under conflicting or concurrent requests, the lack of quantified end-to-end latency characterization for closed-loop agentic workflows, and the missing conformance and certification paths, most acute for LLM xApps operating under sub-10\,ms E2 deadlines, required to transition prototype results into standardized deployment.

\begin{table*}[t]
\centering
\caption{Summary of 6G Project Initiatives Leveraging LLM-Based Agentic AI.}
\label{tab:agentic-6g-projects}
\scriptsize
\rowcolors{2}{gray!10}{white}
\renewcommand{\arraystretch}{1.0}
\setlength{\tabcolsep}{4pt}
\begin{tabular}{p{1.2cm}p{1.3cm}p{1.1cm}p{2cm}p{3.0cm}p{2.5cm}p{2.0cm}}
\hline
\rowcolor{gray!25}
\textbf{Project} &
\textbf{Timeline} &
\textbf{Funding} &
\textbf{Agentic Arch.} &
\textbf{Agentic Contribution} &
\textbf{Integration Point} &
\textbf{Limitation} \\
\hline\hline

SUNRISE-6G &
2024--2027 &
EU SNS-JU &
Single/multi-agent intent mediators over federated testbeds &
LLM agents parse natural-language intents and translate them into platform-specific orchestration primitives via northbound APIs; enables cross-platform service coordination. &
TMF-aligned northbound APIs; federated testbed orchestration layer. &
Intent grounding via prompt engineering; no formal schema alignment across heterogeneous platforms. \\ \hline

6G-INTENSE &
2024--2027 &
EU SNS-JU &
Multi-agent IBN pipeline (spec, decompose, enforce, assure) &
Full intent lifecycle management: LLM agents decompose intents into domain sub-intents, assess real-time feasibility against network state, enforce policies, and close the assurance loop via KPI monitoring. &
ETSI ZSM cross-domain fabric; TM Forum TMF921 intent interfaces. &
No formal convergence guarantees under simultaneous conflicting intents; SLA compliance under KPI degradation unverified. \\ \hline

6G-DALI &
2025--2028 &
EU SNS-JU &
LLM agents as AI lifecycle managers over distributed pipelines &
Agents interpret natural-language intents to trigger DataOps and MLOps workflows: dataset generation, preprocessing, model training scheduling, and experiment reproducibility across NWDAF-compatible nodes. &
NWDAF data pipelines; distributed compute schedulers; model registries. &
End-to-end MLOps loop latency unquantified; consistency under concurrent agent requests on shared infrastructure undefined. \\ \hline

FLECON-6G &
2025--2028 &
EU SNS-JU &
Agentic AI over multi-layer Network Digital Twins &
Agents query digital twin state to compute cross-domain control updates and trigger automated workflows; explainability constraints require causal justification logs traceable to originating intents. &
Multi-layer digital twin control plane; cross-domain intent decomposition with explainable decision logging. &
Twin-reality divergence under anomalies propagates to incorrect agent actions; sync latency under high query rates uncharacterized. \\ \hline

OPEN6G &
2022--Ongoing &
U.S. DoD (IB5G) &
Hierarchical MAS (AgentRAN) across non-RT and near-RT RIC tiers &
LLM rApps consume A1 intent policies at non-RT RIC; lightweight inference xApps act on E2 indications within 10--100\,ms loop; digital twin validates actions before live RAN commitment. &
O-RAN RIC stack; A1/E2 interfaces; programmable Open RAN testbed with digital twin validation. &
No O-RAN conformance procedure for LLM xApps; distilled model accuracy under real traffic at $<$10\,ms E2 deadline undemonstrated. \\

\hline
\end{tabular}
\end{table*}
\rowcolors{0}{}{}

\section{Open Challenges and Future Perspectives}
\label{sec:open-challenges}

While Agentic AI promises to revolutionize telecom network management, it also introduces significant new challenges that must be addressed for 6G.
The convergence of LLM capabilities with telecom requirements creates a unique research frontier where scalability, real-time responsiveness, interpretability, security, and regulatory considerations intersect in unprecedented ways \cite{sec2.b11}. In this section, we discuss the major challenges and open research questions of Agentic AI in telecom networks 5G/6G.

\subsection{Lack of Deterministic Guarantees: Stochasticity, Hallucination, and Cascading Misinformation}
LLMs are inherently stochastic systems whose outputs are sampled from a learned probability distribution over token sequences conditioned on the input context, rather than derived from formally specified operational semantics. This stochasticity is the generative source of the hallucination phenomenon, wherein a model produces syntactically coherent but factually or semantically incorrect outputs with a non-zero and generally unquantifiable probability \cite{10.1145/3703155}. In network management, this property is structurally incompatible with the deterministic guarantees required by telecommunications control planes. A hallucinated root-cause attribution, such as misclassifying a radio-frequency interference event as a hardware failure, can trigger erroneous reconfiguration of live gNodeB parameters with cascading effects across co-scheduled user equipment. The problem is compounded in multi-agent pipelines, where the stochastic output of one LLM agent serves as the grounding context for subsequent agents, enabling what the literature terms cascading hallucinations: misinformation is reinforced through memory retrieval, tool invocation, or inter-agent communication and amplified across multiple decision steps before any corrective mechanism intervenes.

\noindent\textbf{Future Perspectives.} Addressing this challenge requires research into formal uncertainty quantification methods for LLM agents operating in closed-loop network control settings, including Bayesian approaches that produce calibrated confidence estimates alongside network management decisions. Hybrid neuro-symbolic architectures that enforce 3GPP-specified operational constraints as hard logical invariants, rather than as soft learned preferences, offer a promising path toward determinism-compatible agentic control.

\subsection{Computational Cost and Energy Footprint}

The deployment of LLM-based agents as components of network management systems introduces a resource consumption profile that conflicts with the sustainability objectives of 6G, which targets an order-of-magnitude improvement in energy efficiency relative to 5G. State-of-the-art LLMs in the 7B to 70B parameter range require 14 GB to 140 GB of GPU memory for inference, a resource profile that directly conflicts with the 6G target of an order-of-magnitude improvement in energy efficiency relative to 5G. Compression techniques, including 4-bits quantization, structured pruning, and knowledge distillation reduce the memory footprint by factors of four to eight but introduce accuracy degradation on domain-specific reasoning tasks that remains uncharacterized for telecom applications \cite{11383190}. In multi-agent deployments spanning Non-RT and Near-RT RIC layers, NWDAF, and distributed edge orchestrators, aggregate inference energy budgets dominate total system cost: in hybrid LLM-MARL architectures, LLM inference accounts for the majority of system energy expenditure even when invoked in fewer decision cycles.

\noindent\textbf{Future Perspectives.} Purpose-built Small Telecom Language Models targeting 1B to 3B parameters via domain-adaptive fine-tuning, energy-aware agent scheduling policies that gate LLM inference based on decision complexity, and the integration of inference costs as explicit constraints within network resource optimization frameworks are priority directions for future work.

\subsection{Real-Time Inference Latency vs. Network Control Time Scales}
The temporal architecture of 5G/6G control loops imposes strict latency bounds that current LLM inference pipelines cannot satisfy at the functional layers where autonomous control is most consequential. O-RAN Near-RT RIC control loops operate in the 10 ms to 1 s range, whereas LLM autoregressive inference latency is highly dependent on model size, batch size, decoding strategy, and hardware configuration. For instance, large-scale models (e.g., 30B–70B parameters) running with batch size 1 and standard autoregressive decoding (greedy or top-k) on a single high-end GPU (e.g., A100/H100) typically exhibit end-to-end response times on the order of 200 ms to several seconds per query, with per-token latencies in the tens of milliseconds. Smaller models (e.g., 7B–13B) or quantized variants can reduce latency, particularly under larger batch sizes or with aggressive decoding approximations, but this introduces accuracy degradation and still struggles to satisfy strict real-time constraints when accounting for end-to-end pipeline overheads (prompt construction, tool invocation, and network I/O) \cite{pellejero2025agentic}.




\noindent\textbf{Future Perspectives.} Hierarchical control architectures separating LLM strategic planning on second-to-minute horizons from lightweight reactive policies executing on millisecond time scales, combined with formal stability and convergence analysis of LLM-in-the-loop RAN control systems, currently absent from the literature, constitute the most urgent research priorities for this challenge.




\subsection{Security Vulnerabilities in Multi-Agent Systems}
Multi-agent LLM architectures introduce AI-to-AI privilege escalation as a structurally novel vulnerability: an agent that resists direct prompt injection will nonetheless execute identical malicious payloads originating from a peer agent, because current safety training addresses human-to-AI rather than AI-to-AI boundaries. Empirically, 82.4\% of evaluated models are compromised through inter-agent communication, compared to 52.9\% via RAG backdoor attacks and 41.2\% via direct injection \cite{lupinacci2025dark}. In O-RAN, the openness of A1, E2, and O1 interfaces constitutes an ingress vector for adversarial content into xApp input streams, while in the 5G core, a compromised agent interacting with the NEF or  Policy Control Function (PCF) can exfiltrate subscriber data or inject policy rules affecting multiple shared-slice tenants simultaneously.

\noindent\textbf{Future Perspectives.} Zero-trust security architectures for multi-LLM network management systems, enforcing mutual cryptographic authentication of agent identities and continuous behavioral attestation, represent a necessary and underspecified research direction. The development of guardian agent architectures that monitor the chain-of-thought and tool invocation sequences of primary management agents in real time, intervening before policy-violating actions reach NF APIs, merits investigation.

\subsection{Lack of Telecom-Specific Agentic AI Benchmarks}

Although some benchmarks for LLM agents exist (listed in Table \ref{tab:telecom_llm_benchmarks}), they remain limited in scope and do not adequately capture the complexity of telecom network environments. Current benchmarks focus primarily on general reasoning or software tasks, with limited representation of network dynamics, multi-domain interactions, and real-time constraints. The shortage of realistic datasets and evaluation frameworks hinders the ability to assess the performance, robustness, and scalability of agentic AI systems in telecom contexts.

\noindent\textbf{Future Perspectives.} Open benchmark suites covering heterogeneous multi-domain scenarios spanning RAN optimization, core orchestration, fault diagnosis, and intent translation, instantiated on open emulation platforms such as Open5GS, OpenAirInterface, and Free5GC, and incorporating standardized adversarial stress tests referenced to standards like 3GPP and ETSI, are a prerequisite for rigorous scientific progress and operator confidence in agentic network management systems.

\subsection{ Agentic Explainability, Governance, and Regulatory Compliance}
The EU AI Act classifies autonomous systems operating in critical infrastructure as high-risk, mandating transparency, auditability, and human oversight requirements that current LLM-based agentic architectures are structurally unable to satisfy. The autoregressive attention mechanism underlying LLMs does not produce decision traces interpretable in terms of 3GPP parameter bounds, interference constraints, or operator policy rules, and post-hoc explanation methods, rendering them insufficient for regulatory audit purposes \cite{maz2025eurocomply}. In multi-operator and cross-border 6G deployments, agentic decisions mediated by third-party-hosted LLMs raise data sovereignty concerns that prevent operators from exposing the full network state required for effective agent operation without violating national data residency regulations, a structural tension addressed by the Sovereign AI paradigm, which advocates embedding LLM inference within operator-controlled infrastructure under nationally governed AI frameworks, though its implementation within Near-RT and Non-RT RIC components at production scale has not yet been demonstrated.

\noindent\textbf{Future Perspectives.} Future work should prioritize explainable agentic architectures that generate structured, standards-referenced decision logs mapping each management action to specific 3GPP specification clauses, KPI thresholds, and operator policy rules, enabling post-hoc audit trails compatible with regulatory requirements. Governance frameworks incorporating human-in-the-loop validation gates for high-impact control actions, policy-constrained inferencing, and zero-trust model update pipelines are essential operational safeguards. Active collaboration between AI researchers, network operators, 3GPP, ETSI, and national telecommunications regulators will be necessary to establish conformance testing regimes and certification procedures for AI-native NFs prior to large-scale 6G deployment.

\section{Conclusion}
\label{sec:conclusion}

In this paper, we provide a comprehensive, tutorial-based survey of Agentic AI empowered by LLMs and its profound implications for the evolution of 5G/6G Networks. We articulate the progression from early predictive language models to autonomous and collaborative agentic systems, and rigorously examine their integration across 5G infrastructures and emerging visions for 6G. By coherently synthesizing advances in agent-centric intelligence with contemporary network architectures, protocol frameworks, and standardization initiatives, this survey elucidates how Agentic AI enables intent-driven operation, ZSM, and adaptive network orchestration at scale. Beyond consolidating the state of the art, we identify key research frontiers and architectural imperatives, positioning Agentic AI as a foundational pillar for intelligent, resilient, and self-evolving future networks. This work aspires to inform, inspire, and guide both academic inquiry and industrial innovation at the intersection of AI and telecommunications.

\section*{Acknowledgments}
This work is supported by the European Union Horizon Program under the 6G-INTENSE project (Grant No. 101139266) and the FLECON-6G project (Grant No. 101192462).

\bibliographystyle{unsrtnat} 
\bibliography{sample-base}

\end{document}